\documentclass[useAMS,usenatbib]{mn2e}

\usepackage{graphicx} 
      
\def\msun{{\rm M}_\odot}

\def\lsim{\mathrel{\rlap{\lower 3pt \hbox{$\sim$}} \raise 2.0pt \hbox{$<$}}}
\def\gsim{\mathrel{\rlap{\lower 3pt \hbox{$\sim$}} \raise 2.0pt \hbox{$>$}}}

\newcommand{\erisbh}{Eris\textit{BH} }

\title[]{Bar-driven evolution and quenching of spiral galaxies in 
  cosmological simulations}

\author[Spinoso et al.]{Daniele Spinoso$^{1}$\thanks{d.spinoso@campus.unimib.it}, Silvia Bonoli$^{2}$, Massimo Dotti$^{1,3}$, Lucio Mayer$^{4,5}$,
  Piero Madau$^{6}$\and and Jillian Bellovary$^{7}$\\ \\
$^{1}$ Universit\`a degli Studi di Milano-Bicocca, Piazza della
             Scienza 3, 20126 Milano, Italy\\
$^{2}$ Centro de estudios de fis\'i ca del cosmos de Arag\'o n, plaza
			 San Juan, 1 planta-2 44001 Teruel, Spain\\
$^{3}$ INFN, Sezione di Milano-Bicocca, Piazza della
                         Scienza 3, 20126 Milano, Italy \\
$^{4}$ Center for Theoretical Astrophysics and Cosmology, Institute
for Computational Science, University of Zurich,\\ Winterthurerstrasse
190, CH-8057 Z¨urich, Switzerland\\
$^{5}$ Kavli Institute for Theoretical Physics, Kohn Hall, University
of California, Santa Barbara, CA 93106-4030, USA\\
$^{6}$ Department of Astronomy \& Astrophysics, University of California,
1156 High Street, Santa Cruz, CA 95064\\
$^{7}$ Department of Astrophysics, American Museum of Natural History,
Central Park West \& 79th St, New York, NY 10024
}
\begin{document}

\pagerange{\pageref{firstpage}--\pageref{lastpage}} 

\maketitle

\label{firstpage}

\begin{abstract}
\noindent
We analyse the output of the hi-res cosmological ``zoom-in''
hydrodynamical simulation \erisbh to study self-consistently the
formation of a strong stellar bar in a Milky Way-type galaxy and its
effect on the galactic structure as well as on the central gas
distribution and star formation.  The simulation includes radiative
cooling, star formation, SN feedback and a central massive black hole wich is
undergoing gas accretion and is heating the surroundings via thermal AGN
feedback.  A large central region in the \erisbh disk becomes
bar-unstable after $z\sim 1.4$, but a clear bar-like structure starts
to grow significantly only after $z\simeq0.4$, possibly triggered by
the interaction with a massive satellite. At $z\simeq0.1$ the bar
stabilizes and reaches its maximum radial extent of $l\approx2.2$ kpc.
As the bar grows, it becomes prone to buckling instability, which we
quantify based on the anisotropy of the stellar velocity dispersion.
The actual buckling event is observable at $z\simeq 0.1$, resulting in
the formation of a boxy-peanut bulge clearly discernible in the
edge-on view of the galaxy at $z=0$.  The bar in \erisbh does not
dissolve during the formation of the bulge but it is long-lived and is strongly
non-axisymmetric down to the resolution limit of $\sim 100$ pc at
$z=0$. During its early growth, the bar exerts a strong torque on the
gas within its extent and drives gas inflows that enhance the nuclear
star formation on sub-kpc scales. Later on, as the bar reaches
its maximum length and strength, the infalling gas is nearly all
consumed into stars and, to a lesser extent, accreted onto the central
black hole, leaving behind a gas-depleted region within the central $\sim 2$
kpc. Observations would more likely identify a prominent, large-scale
bar at the stage when the galactic central region has already been
quenched. Bar-driven quenching may play an important role in
disk-dominated galaxies at all redshift.  AGN feedback is instrumental
in this scenario not because it directly leads to quenching, but
because it promotes a strong bar by maintaining a flat rotation curve,
suppressing the density of baryons within the central kpc in the early stages of the formation of the galaxy.
\end{abstract}


\section{Introduction}
Bars are extremely common non axisymmetric features in disk galaxies,
occurring in up to $\gsim 30\%$ of massive ($M_* \gsim 10^{9.5} \msun$)
spirals in the local Universe \citep{Laurikainen04,Nair10,Lee12a,gavazzi15}.
Bars are considered to play a key role in the evolution of disk galaxies,
being able to drive strong inflows of gas towards the central galactic regions
\citep[e.g.][]{sand76, Roberts79, Athanassoula92} and triggering nuclear
star-bursts \citep[e.g.][]{Ho97, Martinet97, Hunt99, Laurikainen04,
Jogee05}.
Bars are also thought to be responsible for the build-up
of the pseudo/disky bulges, whose nearly exponential profiles hints to a
disk origin \citep[e.g.][for a review]{kor13}.  These structures are the
most common type of bulges among galaxies in the stellar-mass range $10^{9.5}\msun
<M_*<10^{10.5} \msun$, while classical bulges dominate among more massive
systems \citep[e.g.][]{FisherDrory11}.
Bars can also be responsible for
triggering AGN activity, if a fraction of the inflowing gas can reach the
central sub-pc of the galaxy \citep[e.g.][]{Shlosman89, Berentzen98,
  sellwood99, comb00, quere15, fanali15}.
 
On longer timescales, the removal of the gas forced towards nuclear regions
affects the star formation processes within the bar extent
\citep{cheung13, fanali15}, contributing to the lowering of the specific star
formation rate in the most massive spiral galaxies at low redshift
\citep{cheung13, gavazzi15}. In addition to the effect of the bar onto the
inter stellar medium (ISM), the dynamical evolution of the bar itself is advocated to be responsible
for the boxy/peanut-shaped stellar bulges (B/P bulges hereafter) \citep[see][for a review]{kor13, Sellwood14}, observed in $\gsim 40 
\%$ of 
edge on disk galaxies \citep[e.g.][]{Lutticke00}. Together with the high fraction of disky pseudobulges, the frequent occurrence of B/P bulges hints at the fundamental importance of secular evolution in the shaping of the central regions of disk galaxies.

Most of the theoretical studies that support the existence of causal
connections between bars and the above-mentioned structures/processes are either
analytical or based on  simulations of isolated galaxies
\citep[e.g.][]{Athanassoula92,Berentzen98,Regan04,Berentzen07,VillaVargas10,Kim12,Cole14}. Although
these kind of simulations are extremely informing about the dynamical effect
of bars, cosmological simulations are needed to follow bar formation within
the hierarchical growth of galaxies \citep[as discussed in, e.g.,][]{kor13}.
Furthermore, most of the simulation literature on bar formation and evolution
is based on collisionless simulations. A few works have employed hydrodynamics
and star formation, showing interesting differences on important issues such
as bar survival and bar-buckling \citep[see e.g.][]{deb06, athan13, athan05},
but none of them has employed modern sub-grid recipes for feedback, which constitute
a crucial aspect of recent progress in simulating galaxy formation.

To date only an handful of fully cosmological simulations have achieved the
required numerical resolution and included all the physical processes needed
to self-consistently produce barred galaxies
\citep[e.g.][]{RomanoDiaz08,Scannapieco12,Kraljic12,Goz14,bonol15,
  fiac15, okamoto15}. 
 Among the above-mentioned cosmological simulations
 of barred disk galaxies, \erisbh \citep{bonol15} and Argo \citep{fiac15} share the highest
 spatial and mass resolutions\footnote{The simulations by \cite{Kraljic12} have a comparable
  spatial resolution but a coarser resolution in mass.}, but Argo has been
 evolved only down to $z=3$, while \erisbh has been followed down to $z=0$,
so its properties can be compared directly with the
 observed properties of well-resolved barred galaxies.

 \erisbh is a twin simulation of Eris \citep{gued11}, with which it shares
initial conditions, resolution and sub-grid physics, but, unlike Eris, it also includes
prescriptions for the formation, growth and feedback of supermassive black
holes. Both Eris and \erisbh reseamble, at $z=0$, a late-type galaxy such as the
Milky Way \citep{gued11, bonol15}, but while Eris hosts a typical pseudobulge,
\erisbh features a strong bar and its bulge has a clear
boxy-peanut morphology \citep{bonol15}.

The aim of this work is to study the buildup and the evolution of the strong bar
seen in \erisbh,
to learn about the origin of bars and the impact that these structures have  in
shaping  galaxies like our own Milky
Way.  

The paper is organized as follows. In Section \ref{simulation} we briefly
summarize the properties and main results of the \erisbh simulation. In Section
\ref{sfc_dnst} we study the buildup of the bar, quantifying its strength and
radial extent; we analyze the dynamical properties of the galaxy disk,
testing its stability to non-axisymmetric perturbations and looking
for resonances between the bar bulk precession and the orbital motions of disk
stars; we also analyze the formation of the B/P morphology of the bulge. 
In Section \ref{gas_response} we show the impact of the bar in depleting gas in
and triggering star formation in the central region of the galaxy. 
Finally, in Section \ref{conclusions} we summarize and discuss our results.

\section{The \erisbh simulation}
\label{simulation}

\erisbh \citep{bonol15} is one of the runs in the Eris suite of simulations
\citep[][]{gued11, mayer12, shen12, bird13, shen13, gued13, rashkov13, sokolowska16}
which have been among the first zoom-in cosmological simulations to produce realistic
late-type spirals with properties comparable with the Milky Way at $z=0$. \erisbh \citep{bonol15} inherits its
initial condition and most of its features from the first Eris run
\citep{gued11, gued13}, from which it differs in that it also includes prescriptions for
the formation and accretion of massive black holes (MBHs) and their associated AGN feedback.
Here we summarize
the main characteristics of Eris and the new sub-grid physics implemented in
\erisbh. For more details we refer the reader to \cite{gued11} and
\cite{bonol15}.

Eris was obtained from a zoom-in of a Milky Way-size halo selected within 
a low-resolution, dark matter-only simulation of a $(90\mbox{
  Mpc})^3$ volume. This simulation assumed a flat universe with
$\Omega_{\scriptsize{\mbox{M}}}=0.24$, $\Omega_{\scriptsize{\mbox{b}}}=0.042$,
$h_0=73\mbox{ km s}^{-1}\mbox{Mpc}^{-1}$, $n=1$ and $\sigma_8=0.76$ obtained
from the WMAP three-year data (\cite{sper07}). The target halo was selected
also because of its quiet merger-history
(i.e. no major mergers after $z=3$, where a major merger is defined as an
encounter between two haloes whose mass ratio is above $1:10$). The
cosmological evolution of the haloes was simulated from $z=90$ down to $z=0$
with the parallel N-body spatially and temporally adaptive tree-SPH code
$GASOLINE$ \citet{gasoline}.

Within the high-resolution region, the initial dark-matter and gas particles
masses were set respectively to
$m_{\scriptsize{\mbox{DM}}}=9.8\times10^4\mbox{M}_\odot$ and
$m_{\scriptsize{\mbox{g}}}=2\times10^4\mbox{M}_\odot$.
The gravitational
softening length was fixed to the value of $\varepsilon_0=120$ physical parsec
for each particle type from $z=0$ to $z=9$ and evolved as
$\varepsilon(z)=\varepsilon_0(1+z)^{-1}$ from $z=9$ to $z=90$. \erisbh, as the 
original Eris, includes
recipes for Compton and atomic cooling, heating from a UV background and
radiative cooling. Energy and metals injection in the interstellar medium due
to SNe explosions and stellar feedback are modelled following the recipe of
\cite{stin10}. No metal-line cooling or metal diffusion were included.

Owing to the high resolution of the simulation we could use a relatively high
density-threshold for star formation,
i.e. $n_{\scriptsize{\mbox{SF}}}=5$ atoms/cm$^3$. The combination of SNe
feedback and the high density-threshold for star formation produces a realistic
clumpy interstellar medium \citep{gued11} and
removes low-angular momentum gas from the simulated disk. The final outcome of
\erisbh is a Milky Way-size disk galaxy with a low bulge-to-disk (B/D) ratio
and a flat rotation curve (with rotation velocity at the solar radius of
$190\pm15$ km/s),  whose location on the Tully-Fisher,
stellar-mass/halo-mass, and stellar velocity dispersion-MBH mass  relations is consistent with that of the
Milky-Way.
Note that at $z > 0.5$ both \erisbh and Eris have overly
efficient star formation relative to the abundance matching predictions, while
they agree with it at $z=0$. Recent runs in the Eris suite which incorporate
both metal-line cooling and stronger SN feedback, do obey abundance matching
constrainst at higher redshift but miss a kinematically cold thin disk component
at $z=0$ (\cite{sokolowska16}; Mayer et al., in preparation).

\begin{figure*}
\centering
\includegraphics[width=0.9\textwidth]{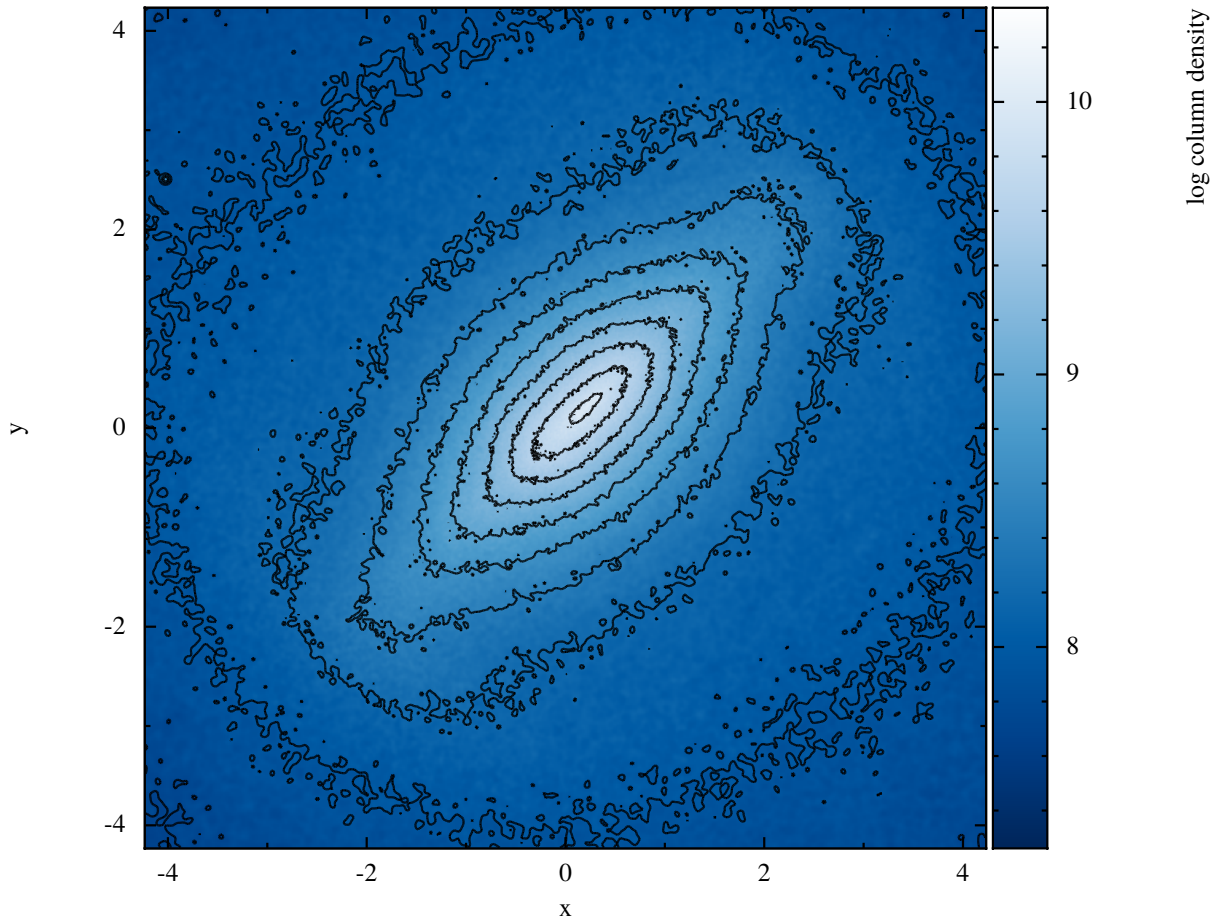}
\caption{Stellar surface density $\Sigma_{*}$ map of the central region of the
 \erisbh 
  galaxy at redshift $z=0$. The $x$ and $y$ axis units are in [kpc h$^{-1}$]
while the color shades
  show the values of $\log(\Sigma_*/M_\odot \mbox{kpc}^{-2})$. The black solid
  lines are the iso-density contours used to reveal the inner structure of the
  bar and to show that the non-axisymmetric shape is maintained even at very
  small radius. Contours are separated by a $0.2$ difference in
  $\log(\Sigma_*)$ starting from a value of $\log(\Sigma_*/M_\odot
  \mbox{kpc}^{-2})=9.8$ in the centre. Note that the deviation from
  axisymmetry increases at smaller and smaller radii so that the central bar
  structure is more elongated than the global bar structure. This feature is
  common in all the snapshots where the bar is clearly visible.}
\label{render}
\end{figure*}

\erisbh includes recipes for the seeding,
growth and thermal feedback of MBHs.
Growth occurs by both mergers with
other MBHs and gas accretion.
All the other parameters in
this new run were kept identical to those in the original Eris in order to allow a coherent
comparison between the two simulations. In \erisbh, a MBH seed is placed in
every halo that $(i)$ does not already host a MBH, $(ii)$ is resolved
with at least $10^5$ particles, and $(iii)$ hosts at least $10$ gas particles in
regions denser than $100$ atoms/cc. Only four proto-galaxies in the simulation match the
abovementioned conditions before $z\sim3$ and are thus seeded with a MBH, whose mass is proportional
to the size of the high-density gas region. After $z\sim3$ the gas
  density becomes generally too low for the seeding process to
  occur \citep[][]{bonol15}, The four black hole seeds are then
allowed to accrete mass following the Bondy-Hoyle-Lyttleton
prescription capped at the Eddington limit, as implemented in
\cite{bellovary10}.  During the accretion phase it is assumed that a
small fraction $\epsilon_{\scriptsize{\mbox{f}}}=0.05$ of the total AGN
luminosity couples with the surrounding gas and heats it. 
The growth of the
black hole hosted by the central galaxy is mostly due to mergers with  black
holes hosted by infalling satellite galaxies, while growth by gas accretion is
very modest,  as reflected by the low accretion rates measured, typically between $10 ^{-3}$ and $10
^{-5}\,\mbox{M}_\odot$/yr (i.e. only $10^{-2} - 10^{-4}$ of the Eddington
limit) \citep[an exhaustive and
extended discussion on the growth of the black holes in Eris{\rm BH} can be found in][]{bonol15}. 
Despite the limited gas growth, the modest feedback energy released by the
central black hole  still manages to affect the
large-scale properties of the host galaxy. For example, \erisbh features a smaller bulge and a more
extended disk when compared to Eris. Because of the absence of a prominent
central mass concentration, the disk in \erisbh is prone to dynamical 
instabilities during late evolutionary stages \citep[e.g.][and references
  therein]{kor13} and a clear stellar bar develops within the central $\sim$3
kpc of the disk. A qualitative analysis of the stellar surface density field in late
  evolutionary stages of \erisbh can easily point out the presence of a
central non-axisymmetric feature, i.e. a stellar bar \citep[see the lower-left
  panel of figure 10 in][and Figure~\ref{render} below for a zoom-in of the
  central galactic regions]{bonol15}.

In the next sections we focus on studying the properties of such bar and its
effect on the host galaxy.


\section{BAR FORMATION AND EVOLUTION}
\label{sfc_dnst}

In this section we first focus on the analysis of the buildup of the bar of
\erisbh, by quantifying its strength and spatial extent across time. We then
study the dynamical stability of the galactic disk, to determine the conditions
that led to the developement of the bar. Finally, we study the emergence of the
B/P morphology of the bulge and connect it to the growth of the bar.

\subsection{Properties of the bar}\label{sec:bar_strength}

In order to quantitatively assess the bar extent and strength we perform a
Fourier decomposition of the projected stellar density field $\Sigma_*(x;y)$ on the disk
plane,
%
by calculating the \textit{cumulative} $A_2$ amplitude
\citep[as in][]{dub09, fiac15}:
\begin{equation}
A_2(r)=\frac{1}{M}\sum_{j=1}^{N} m_j\,\,e^{2i\,\phi_j}\,,
\label{fourint}
\end{equation}
where the summation is carried over the entire set of $N=N(r)$ star particles
up to a distance $r$ from the centre and $M$ is the total mass
within the same distance.  Due to its definition, $A_2(r)$ increases up
to the distance at which the $\Sigma(x;y)$ field structure exhibits a strong
non-axisymmetric component and then gradually falls to zero. The radial
position $r_{\scriptsize{\mbox{A2}}}$ of the maximum value
m$_{\scriptsize{\mbox{A2}}}=\small{\mbox{max}}[A_2(r)]$ is used as an estimate
of the bar radial extent. At the same time, the value of
m$_{\scriptsize{\mbox{A2}}}$ itself can be used as an estimate of the bar
strength, as it measures the bar intensity with respect to the mean projected
density field up to $r=r_{\scriptsize{\mbox{A2}}}$.

\begin{figure}
\centering
\includegraphics[width=0.51\textwidth]{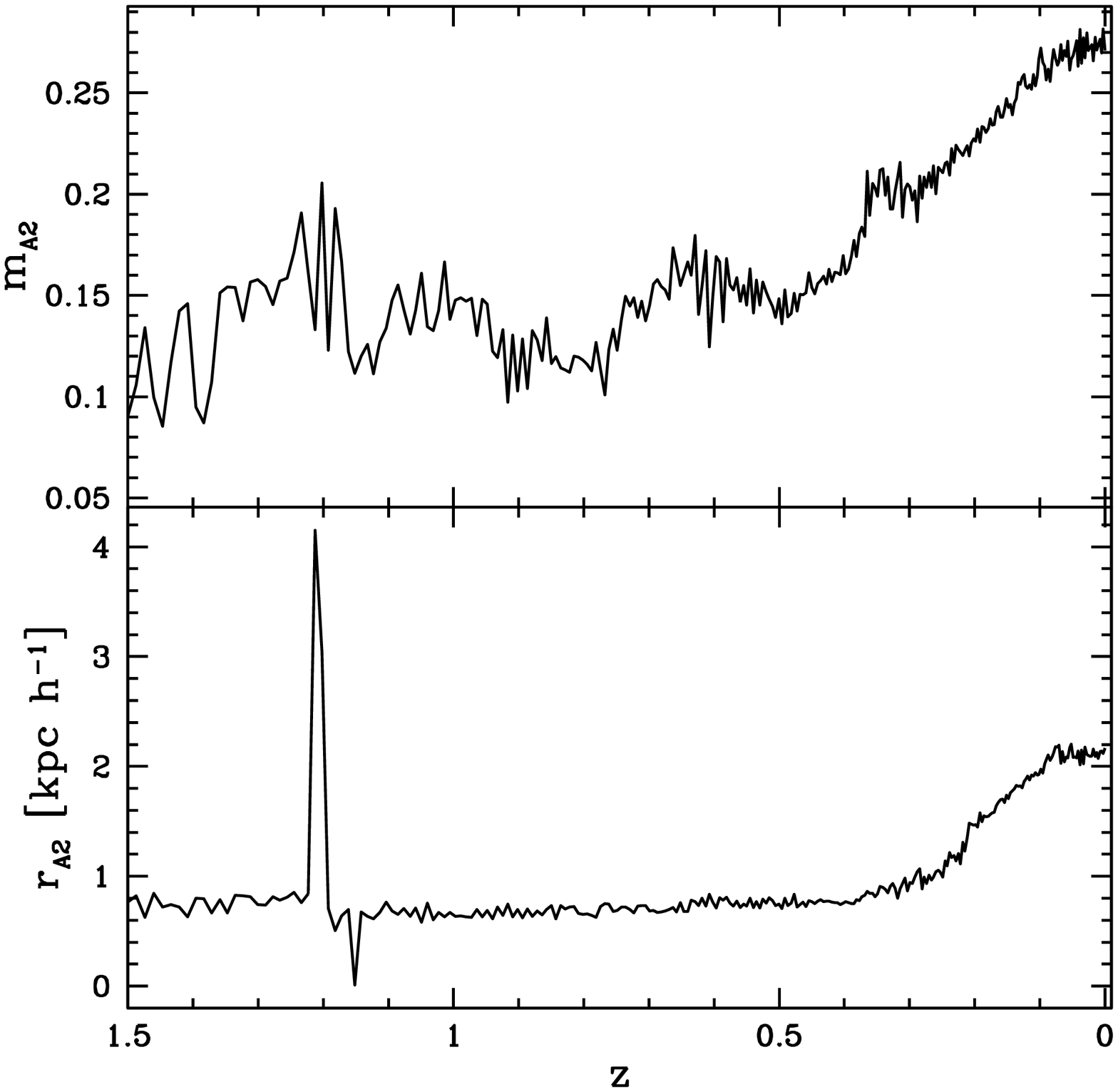}
\caption{Evolution with redshift of m$_{\scriptsize{\mbox{A2}}}$ (upper panel)
  and $r_{\scriptsize{\mbox{A2}}}$ (lower panel). The early fluctuation at
  $z\approx 1.2$ is caused by the last minor merger experienced by the main
  galaxy. A clear transition towards constantly increasing values of
  m$_{\scriptsize{\mbox{A2}}}$ and $r_{\scriptsize{\mbox{A2}}}$ is observable
  at $z \lsim 0.5$, associated with the growth of the galactic bar. A
  flattening in the m$_{\scriptsize{\mbox{A2}}}$ and
  $r_{\scriptsize{\mbox{A2}}}$ profiles is observable at low redshift $z\lsim
  0.1$ in correspondance of the boxy-peanut bulge formation, as discussed in
  the following.}
\label{a2r2fig}
\end{figure}

We calculate the $A_2(r)$ radial profile at each snapshot in order to trace
the bar amplitude evolution as well as its radial extent evolution through
time (Figure \ref{a2r2fig}). During the early stages of disk formation
strong fluctuations in $m_{\scriptsize{\mbox{A2}}}$ are due to ongoing minor
merger events and/or the associated galaxy relaxation events. The last minor
merger occurs at $z \sim 1.2$, after which the galaxy evolves pratically in isolation.

From $z\sim 0.5$ and onwards, the intensity of $m_{\scriptsize{\mbox{A2}}}$ gradually
increases with time and reaches its maximum $m_{\scriptsize{\mbox{A2}}}
\approx 0.27$ close to the end of the simulation. Results in Figure
\ref{a2r2fig} show that the bar radial extent reaches its maximum value
$r_{\scriptsize{\,\mbox{max}}}\approx 2.2$ kpc at late simulation
stages. The bar extent stabilizes about $r\approx2.1$ kpc after $z\lsim 0.1$, in correspondence
with the growth of a central B/P bulge (see below).
The bar strength and the formation time we measure in ErisBH are
consistent with previously published results obtained from both isolated and
cosmological simulations \citep[e.g.][]{Kraljic12, Cole14, fiac15, polyachenko16}, although we
note that a large scatter in particular in the growth time (from $\lsim 1$
Gyr to $\gsim$ 3 Gyr for Milky Way like galaxies) is present in literature.


\subsection{Dynamical stability of the galactic disk}

\begin{figure*}
\centering
\includegraphics[width=0.3\textwidth]{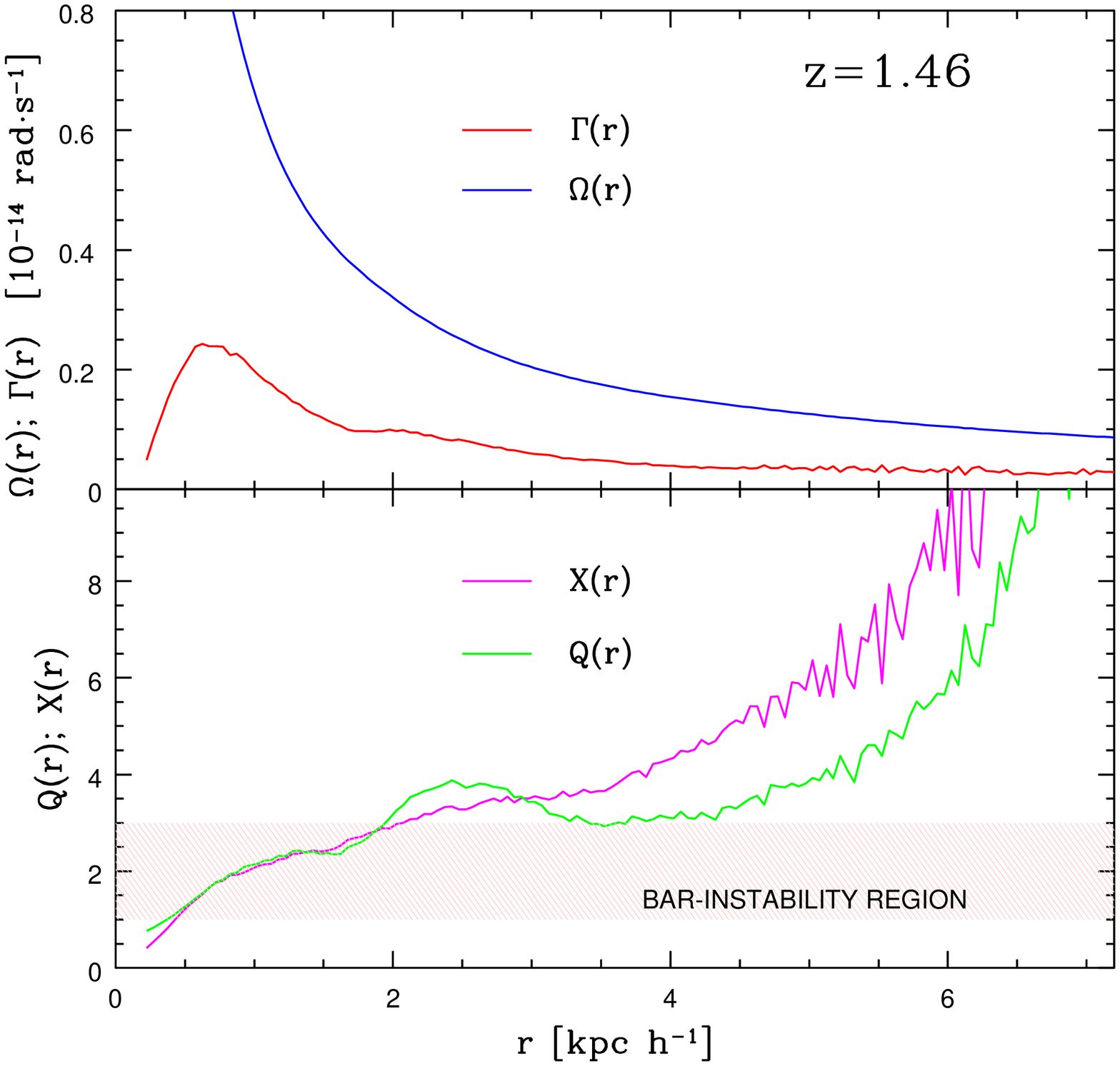}
\includegraphics[width=0.36\textwidth]{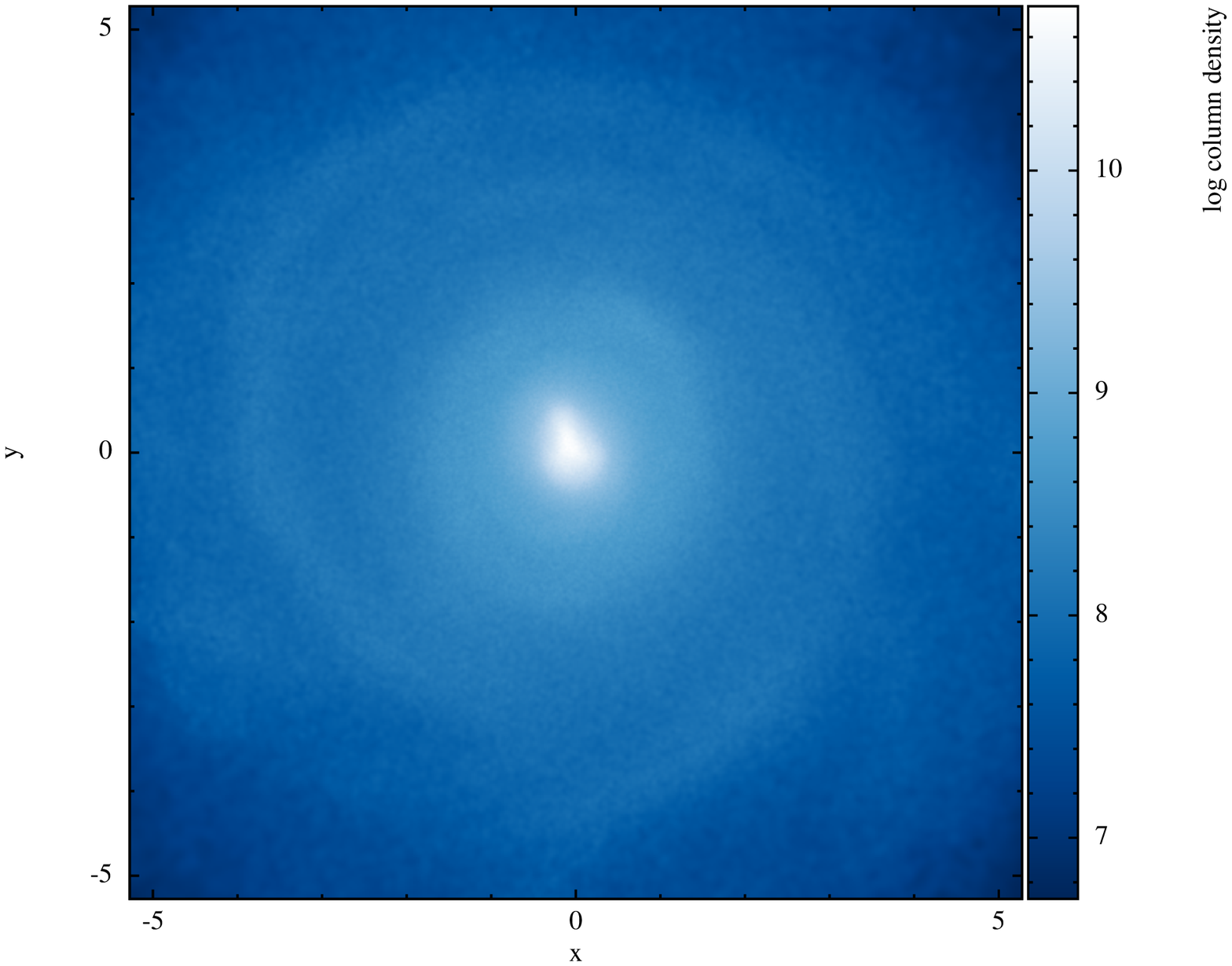}\\
\includegraphics[width=0.3\textwidth]{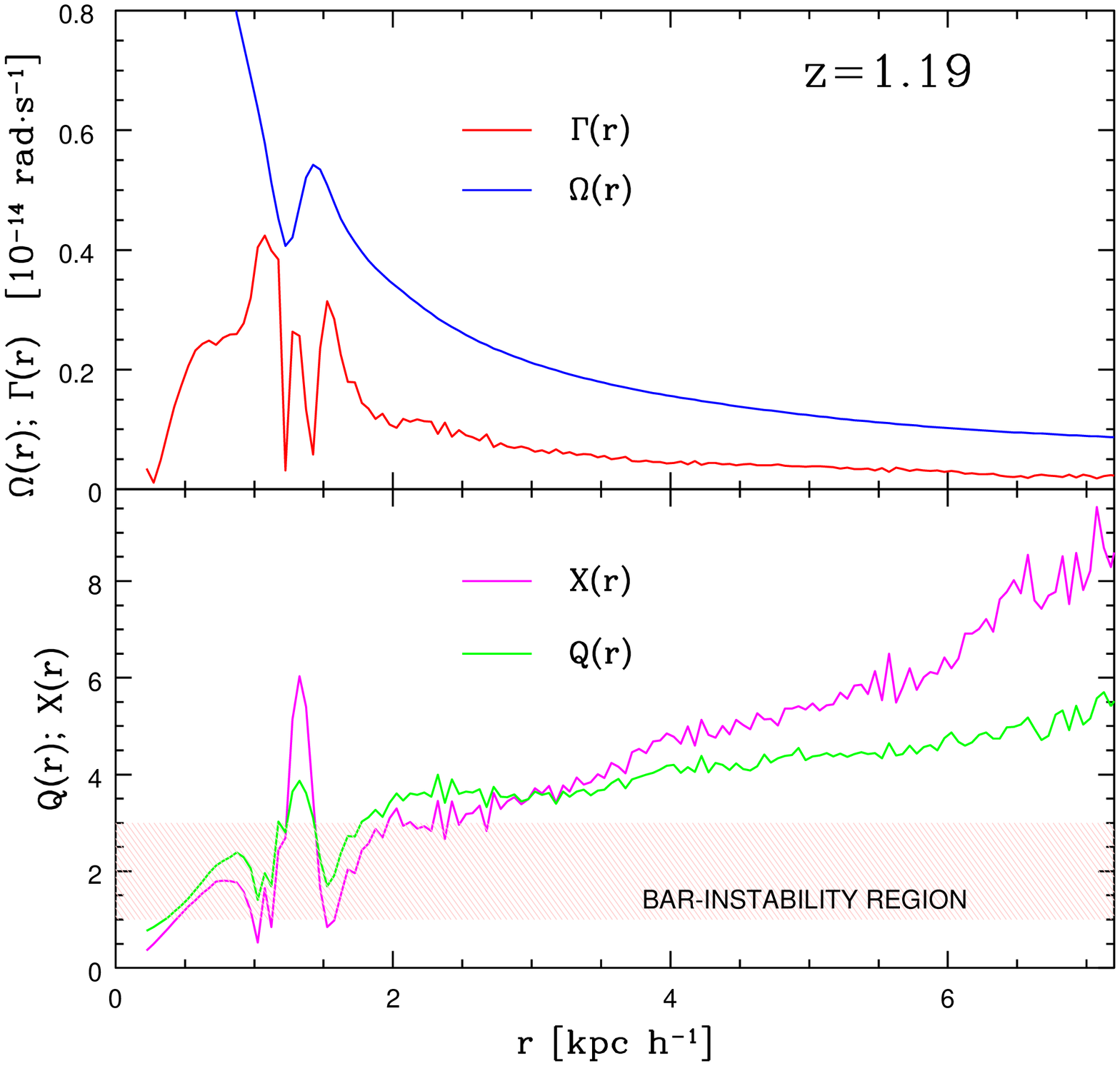}
\includegraphics[width=0.36\textwidth]{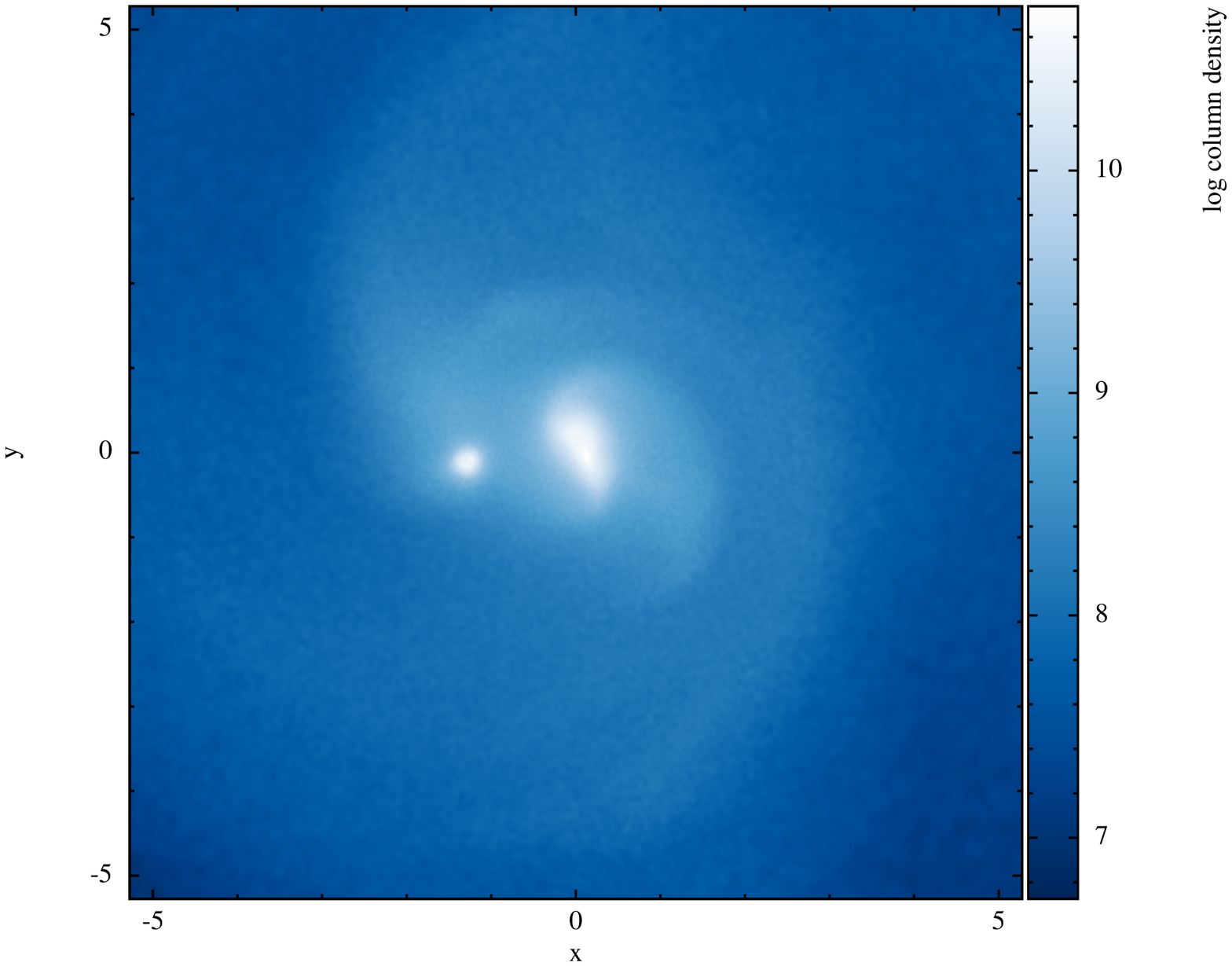}\\
\includegraphics[width=0.3\textwidth]{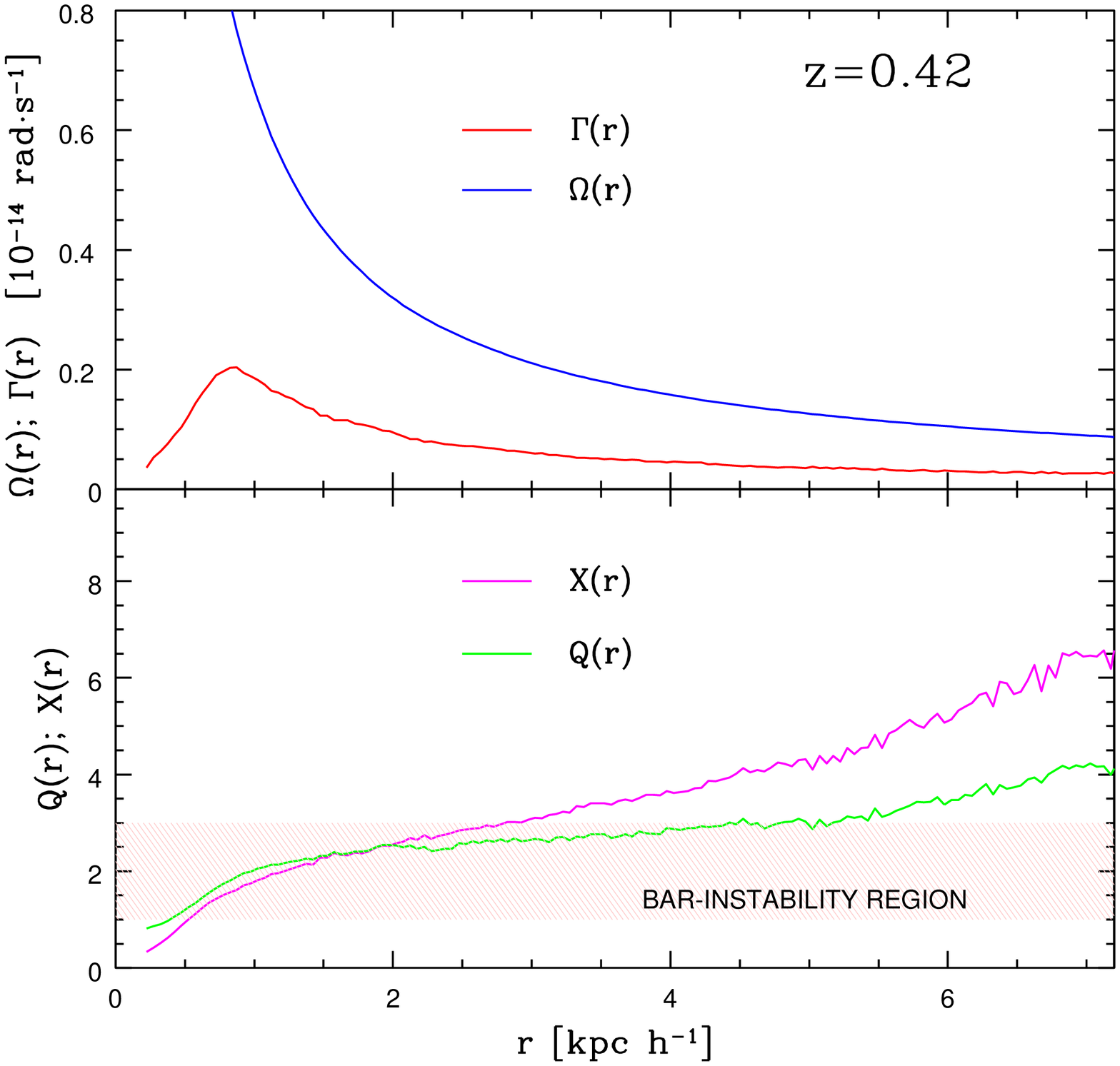}
\includegraphics[width=0.36\textwidth]{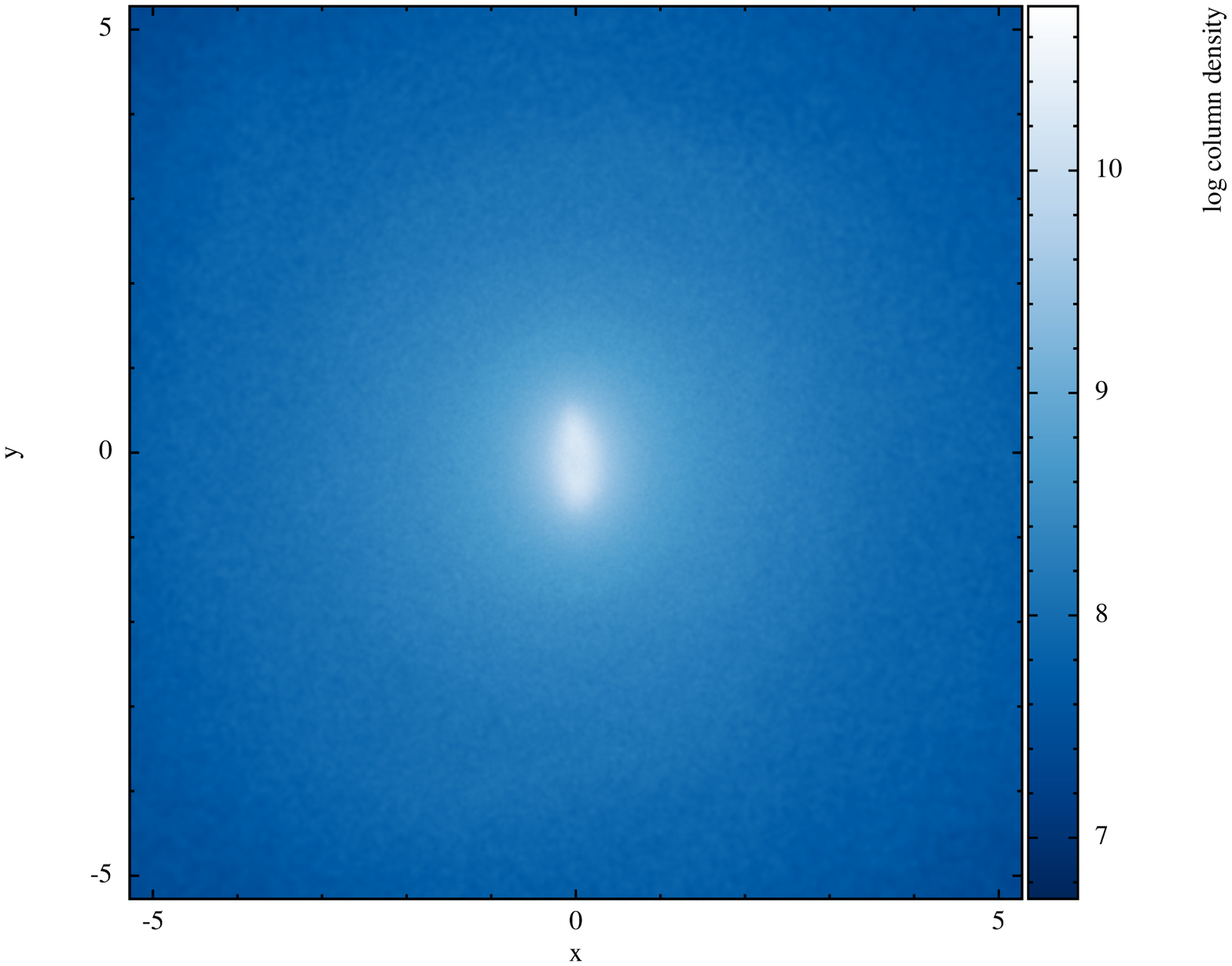}\\
\includegraphics[width=0.3\textwidth]{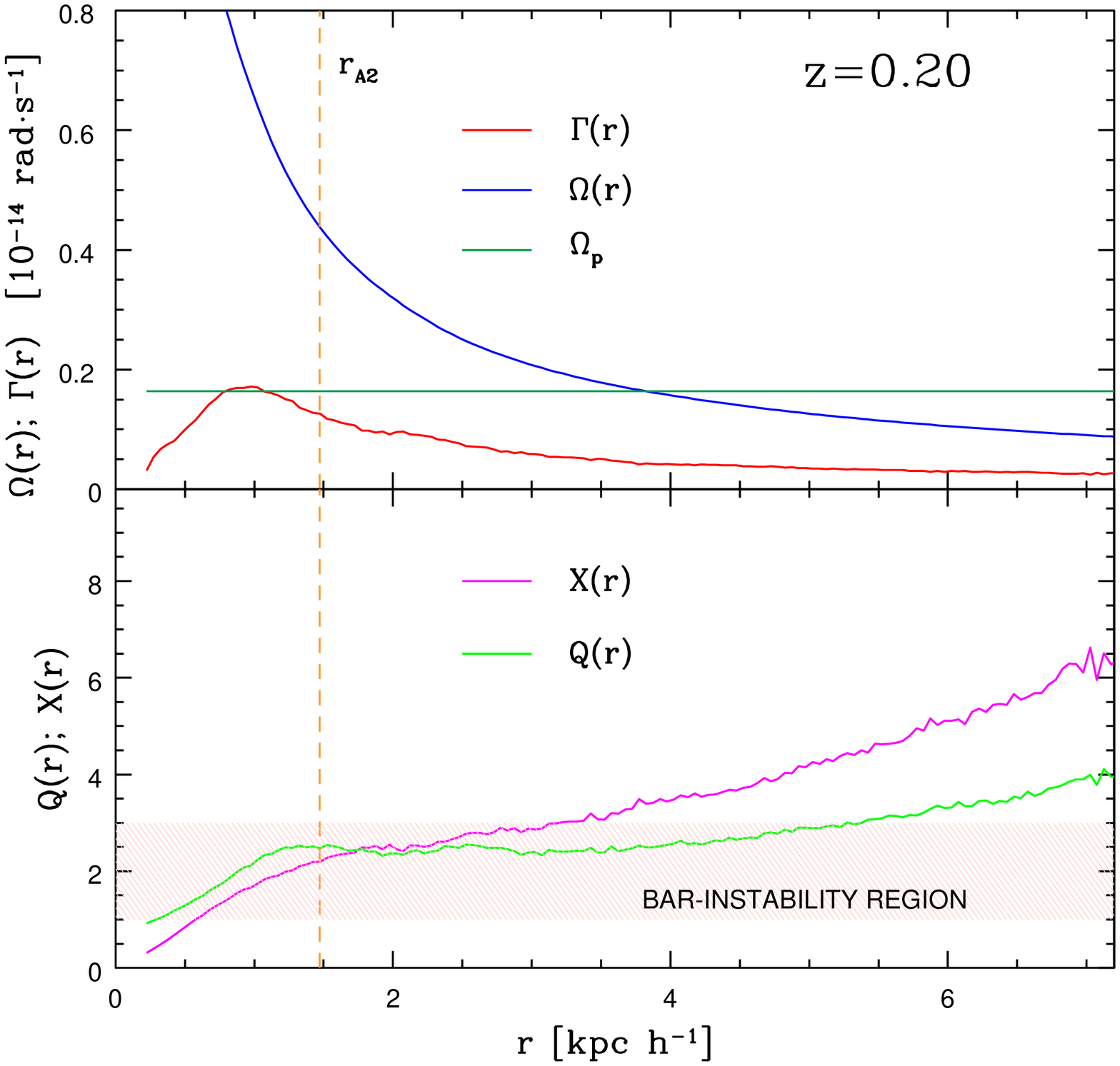}
\includegraphics[width=0.36\textwidth]{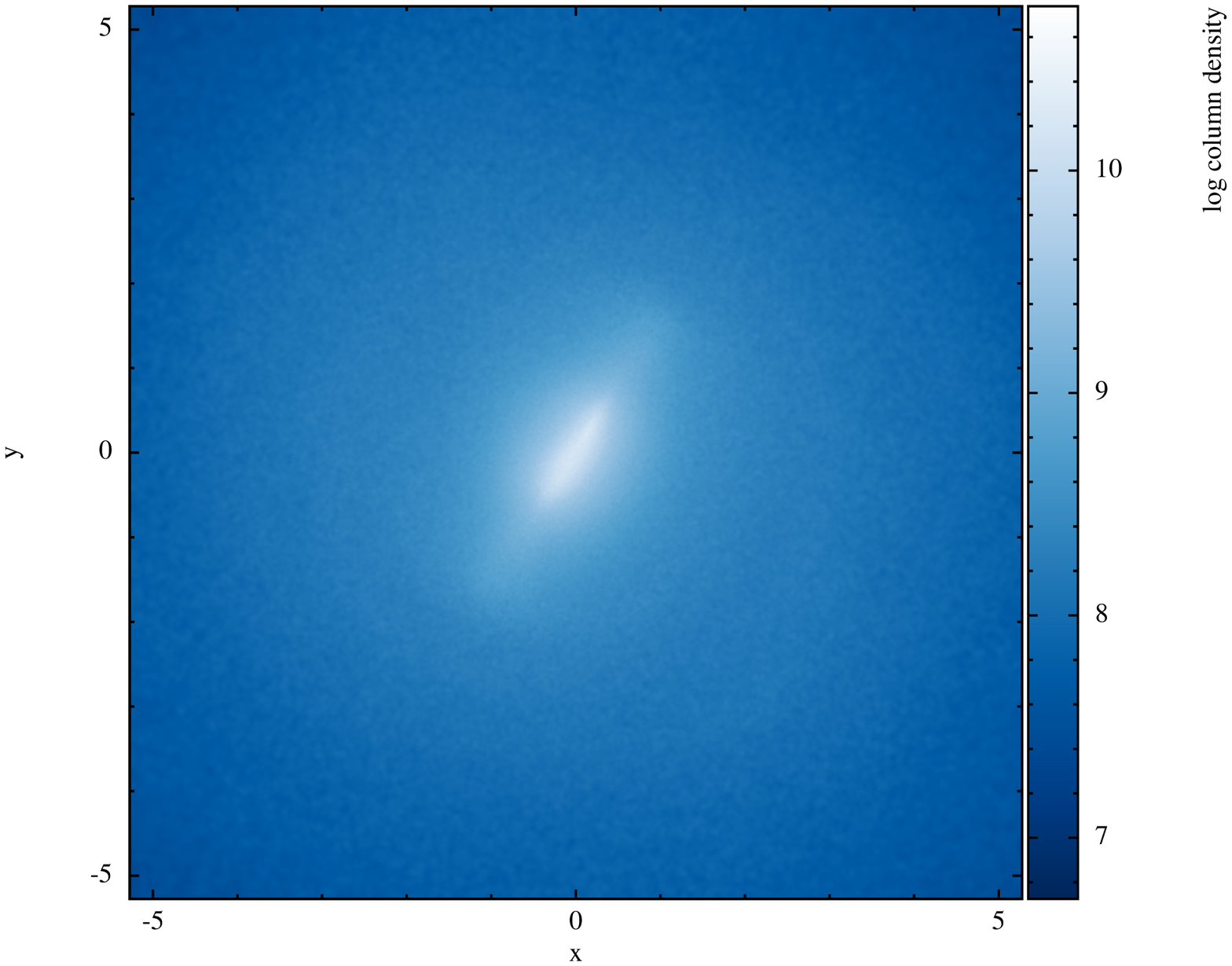}
\caption{Dynamical and morphological structure of the main galaxies at
  different evolutionary stages. From top to bottom: before ($z=1.46$), during
  ($z=1.19$) and after ($z=0.42$ and $z=0.20$) the occurrence of the last
  minor merger. Left panels: frequency plots ($\Omega$ and $\Gamma$, upper
  half) and $Q$ and $X$ stability parameters as a function of the radius.
  The red-shaded area highlights the bar instability region
  ($1\lsim Q\lsim2$ and $1\lsim X\lsim 3$) in each panel. The orange vertical dashed line
  in the bottom panel marks the bar extent, while the
  horizontal green line refers to the bar rotation frequency. Right panels:
  face-on projection of the stellar density map at the corresponding
  redshifts.  Colors encode the stellar surface density (in units of M$_\odot$
  kpc$^{-2}$) on a logarithmic scale. The merging companion is clearly visible
  in the second panel from the top.}
\label{growth}
\end{figure*}

The absence of a central massive bulge makes the galaxy naturally unstable to
the growth of a bar as soon as it settles in a dynamically cold rotationally
supported structure.
For $z< 1.5$ the disk dynamical properties
  allow for the amplification of density perturbations through the
  \textit{swing amplification effect} \citep[see e.g.][]{binney} which may
  easily promote the growth of a bar-like structure.  The effectiveness of this
  process is linked to both the Toomre parameter $Q$ and the swing
  amplification parameter $X$ \citep[see e.g.][]{Toomre64, Goldreich78,
    Goldreich79}. For a differentially-rotating stellar disk the two
  parameters are defined as \citep[see e.g.][]{binney}
\begin{equation}
Q(R)=\frac{\sigma_r(R)\\,\kappa(R)}{3.36\,G\Sigma(R)}\ ;\quad X(R)=\frac{R\,\,\kappa^2\!(R)}{4\,\pi\,G\Sigma(R)}
\end{equation}
where $\sigma_r(R)$ is the radial velocity dispersion of the stars,
$\kappa(R)$ is the epicyclic frequency, $G$ is the gravitational constant and
$\Sigma(R)$ is the star surface density. The Toomre parameter accounts for the
disk stability to axisymmetric density perturbations: if $Q\leq1$ the disk is
unstable. On the other hand the swing amplification parameter $X$ quantifies whether non-axisymmetric perturbations
can grow. Two conditions must be simultaneously verified for the swing
amplification to be effective: $Q\gsim1$ so that the disk is stable but still
strongly responsive to density perturbations, and $X\lsim3$ to prevent the
density waves from being too tightly wound \citep[see][]{binney}.

Figure~\ref{growth} shows the $Q$ and $X$ radial profiles
calculated at four different times. As the two
parameters are in the range $1\lsim Q\lsim2$ and $1\lsim X\lsim3$ (figure~\ref{growth}, red-shaded
areas in left panels), it is clear that an
extended central region (i.e. up to $r\sim3$ kpc) is prone to bar instability.
For reference, the face-on view of the stellar surface density map of the
galaxy is shown in the right panels. The effect of the minor merger happening
at $z\approx 1.2$ on stellar dynamics is clearly observable both in the
$Q$ and $X$ profiles (that show local peaks at the location of the satellite),
as well as in the frequency plot, showing both the angular velocity $\Omega$
and the precessional frequency $\Gamma=\Omega-\kappa/2$, where $\kappa$ is the
epicyclic frequency. This merger imprints a degree of non-axisymmetry on the
central stellar distribution. It is however unclear whether the merger-driven
asymmetric structure is the seed of the stellar bar observable at lower
redshifts or not. Because of the noisy evolution of the
$m_{\scriptsize{\mbox{A2}}}$ parameter at $z\gsim 0.5$ it is impossible to
firmly assert that the bar starts growing already at $z\approx 0.8$ ($1.5$ Gyr
after the completion of the merger) or only at $z\approx 0.5$ (about $3.5$ Gyr
after the merger). For such reason we refrain from commenting further on the
trigger of the bar instability in this section. A discussion about possible
future investigations designed to answer this particular question is presented
in the conclusions. 

The angular frequency ($\Omega_{\rm bar}$) and extent of the bar are shown in
the lower left panel of Figure~\ref{growth} (horizontal green and vertical red
lines, respectively). The bar rotates with a frequency of
$\approx 30$ km s$^{-1}$ kpc$^{-1}$ at $z=0$ wich is similar to the frequency estimated for the Milky Way \citep{gerhard11} and approximately equal to the maximum of
$\Gamma(R)$. This is somewhat expected, since perturbations with $\Omega_{\rm
  bar} \approx {\rm max}(\Gamma)$ are the fastest to grow, as demonstrated for
the first time by \cite{sand77}. The lack of a clear Inner Lindblad resonance
(ILR, defined by the equivalence $\Omega_{\rm bar} =\Gamma(R_{\rm ILR})$), of
the kind of those observable in presence of a strong central concentration of
matter (where $\Gamma$ tends to diverge for small radii) maintains the
elongated bar like structure even at small (sub-kpc) radii (see
figure~\ref{render}). The consequences of the absence of a clear ILRs on the
fate of the bar-perturbed gas will be discussed in the next section.

It is also evident that the bar does not extend out to its corotational radius
($R_{\rm cor}$ defined by the $\Omega_{\rm bar}=\Omega(R)$ equality), but
stops at considerably smaller radii ($r_{\scriptsize{\mbox{A2}}}\sim 0.5
R_{\rm cor}$), in agreement with the results of previously published
cosmological \citep[e.g.][]{okamoto15} as well as in idealized simulations
of tidally induced bars
\citep[e.g.][]{lokas16}. We stress however that the
$r_{\scriptsize{\mbox{A2}}}/R_{\rm cor}$ ratio we found is considerably
smaller than that of most of the observed bars\citep[e.g.][and references
  therein]{aguerri15}, although some galaxies host bars whose
$r_{\scriptsize{\mbox{A2}}}/R_{\rm cor}$ ratios are consistent with the ones
we find \citep{rautiainen08}. On a theoretical ground, small
$r_{\scriptsize{\mbox{A2}}}/R_{\rm cor}$ ratios have been predicted both for
bars triggered by interactions \citep{miwa98} possibly like the one discussed
here and for bars growing in galaxies with an
initially low bulge to disk mass ratio \citep{combes93}, as it is the case of 
the \erisbh simulation.

\subsection{The emergence of the B/P morphology of the bulge}\label{sec:bp}

\begin{figure}
\centering
\includegraphics[width=0.51\textwidth]{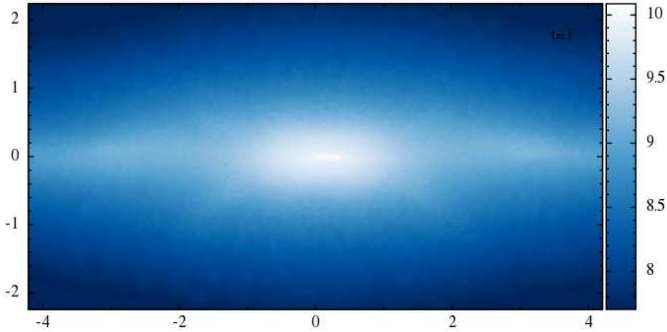}
\caption{Edge-on view of the Eris\textit{BH} last snapshot (redshift $z=0$) in which the
  boxy-peanut shape is clearly visible.  The bar major axis is perpendicular
  to the line of sight to enhance the visibility of the boxy-peanut
  structure. Units are the same as in figures~\ref{render} and \ref{growth}.}
\label{boxpean}
\end{figure}

\begin{figure}
\centering
\includegraphics[width=0.49\textwidth]{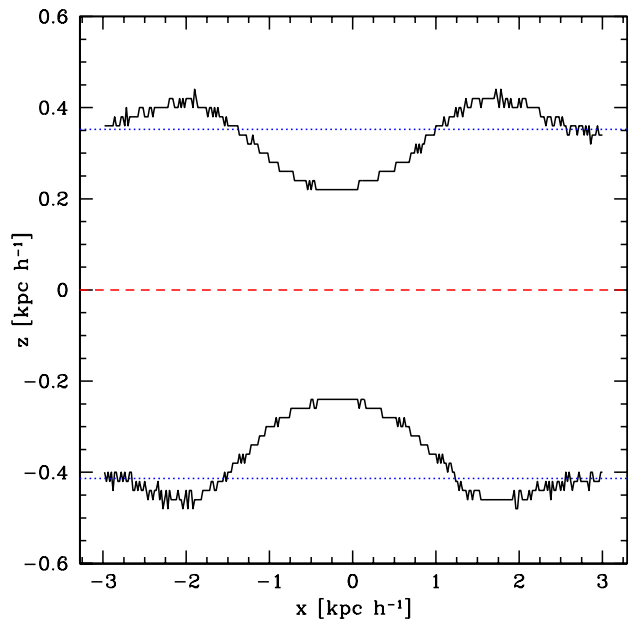}
\caption{$|z|^+$ and $|z|^-$ profiles (red dashed lines) with respect to the
  $x$ coordinate, computed above and under the disk plane, respectively, at redshift $z=0$.
  A double-horned feature is evident in both profiles, demonstrating the
  presence of a boxy-peanut structure in the central region of the galaxy
  \citep{ian15}. The dashed horizontal red line marks the position of the
  galactic plane in the $(x;z)$ plane. The blue dotted lines are reference
  lines used to calculate the relative intensity of the peaks in the $|z|^+$
  and $|z|^-$ profiles (see text).
}
\label{zmed}
\end{figure}

As already commented, the bar stops growing when a B/P structure starts to
form in the central region of the disk. The B/P feature
can be easily pointed out by a qualitative edge-on view analysis of the
Eris\textit{BH} latest evolutionary stages \citep[see][and Figure
  \ref{boxpean}]{bonol15}. 
To constrain the time evolution of the B/P structure we perform a
quantitative analysis on the edge-on projected density field at each
snapshot. We first select the ($x;y$) plane defined by the bar major axis and
the direction perpendicular to the disk plane. On such plane we measure the
$|z|^+(x)$ and $|z|^-(x)$ locations of the median value of the $\Sigma_*$
above or below the disk plane as a function of the $x$ position \citep[as
  in][]{ian15}.  Figure \ref{zmed} shows an example of the $|z|^\pm(x)$
behaviour with respect to the $x$ coordinate at redshift $z=0$. 

\begin{figure}
\centering
\includegraphics[width=0.49\textwidth]{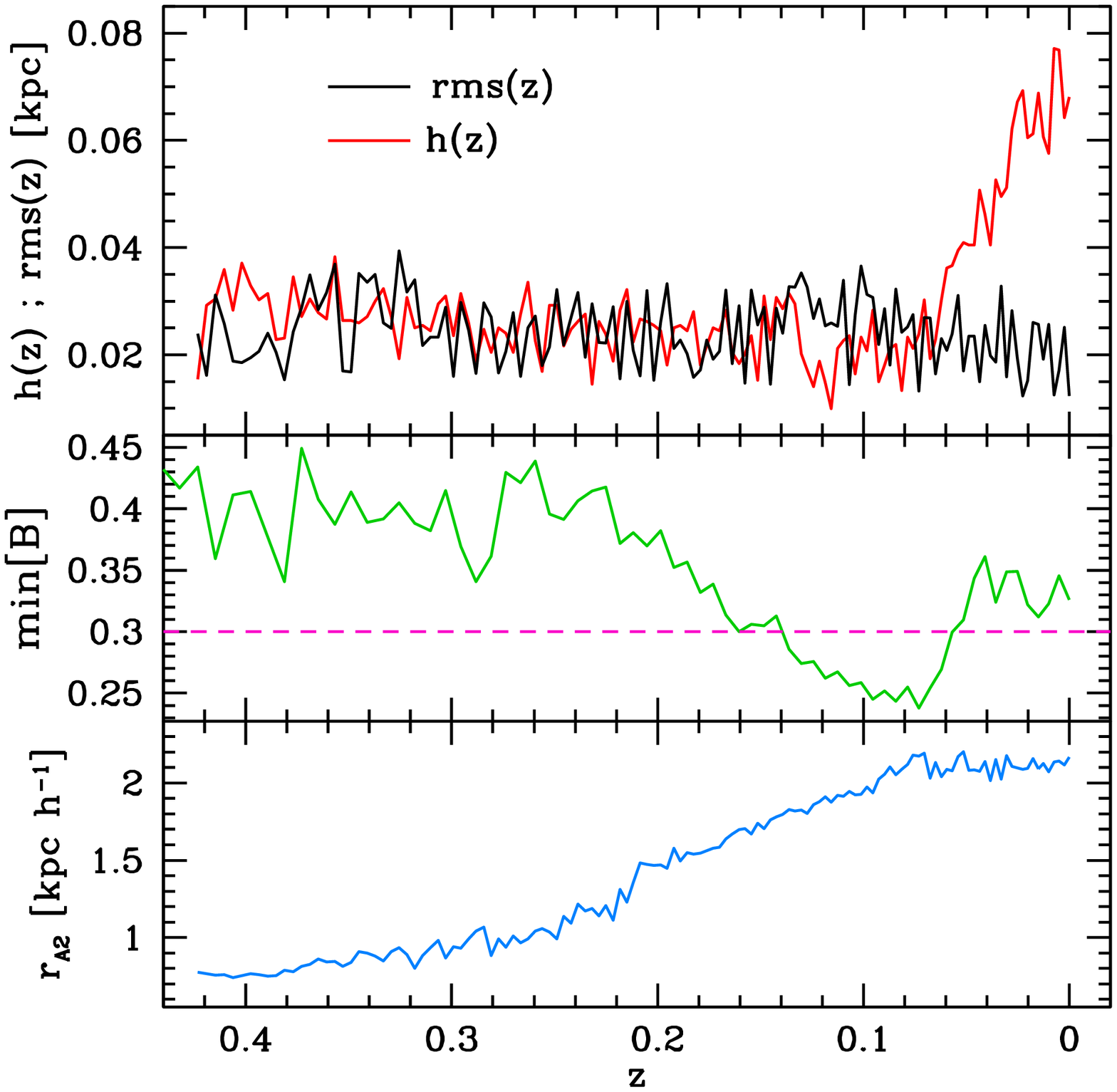}
\caption{Upper panel: B/P strength as a function of redshift $z$. The red
  line refers to the relative height $h$ of the $|z|(x)$ peak to a reference
  value $z_0$, while the error on the $z_0$ average $\sigma_{\!r}$ is
  shown in black. Middle panel: minimim value of the parameter
  $\rm B=(\sigma_z/\sigma_x)^2$ within the inner 3 kpc from the main galaxy
  centre, as a function of redshift. The bar is buckling unstable for
  $\rm B\lsim 0.3$. Lower panel: evolution of the bar length
  $r_{\scriptsize{\mbox{A2}}}$ in the same redshift interval. Note that $h(z)$
  becomes consistently bigger than $\sigma_{\!r}$ towards the end of the
  simulation (upper panel), when the bar stops increasing its size and strength (lower panel).}
\label{boxin}
\end{figure}

A double-horned shape is clearly observable in the $|z|^+(x)$ and $|z|^-(x)$
profiles. To study the growth in time of the B/P
bulge, we first calculate two reference values $z_0^+$ and $z_0^-$ on the
$|z|^+$ and $|z|^-$ profiles respectively, by averaging $|z|^+$ and
$|z|^-$ in the intervals $x\in[-4;-3]$ and $x\in[3;4]$ (outside the bar
region, in the unperturbed disk, see the blue dotted lines in Figure
\ref{zmed}).  This reference value is then used to measure the quantity
\begin{equation}
 h=\mbox{max}\!\left[\,\left|z\right|\,\right]-z_0
\end{equation}
on the four quadrants of the disk projection, and the average of the four
values $h_{\scriptsize{\mbox{m}}}$ is compared with
$\sigma_{\!r}=\small{\mbox{max}}[\sigma^+;\sigma^-]$ where $\sigma^+$ and
$\sigma^-$ are the standard deviations of the $|z|^+$ and $|z|^-$ profiles around the
reference values $z_0^+$ and $z_0^-$ in the outer disk\footnote{We take the
  maximum between $\sigma^+$ and $\sigma^-$ to be more conservative}. If no
double-horned feature is present in the $|z|^\pm(x)$ profiles, then
$h_{\scriptsize{\mbox{m}}}$ must be comparable to $\sigma_{\!r}$.
The results of this analysis are shown in the upper panel of
Figure~\ref{boxin}. 
Clearly $h_{\scriptsize{\mbox{m}}}(z)$ becomes consistently bigger than
$\sigma_{\!r}$ only after redshift $z\simeq0.1$, i.e. the double-horned
feature (and so the B/P structure in the bulge) develops at late evolutionary
stages, when the bar is already strong, as shown by the
$r_{\scriptsize{\mbox{A2}}}$ evolution (shown for $z\lsim 0.4$ in the lower
panel of Figure \ref{boxin} for any easy comparison). In order to constrain
the origin of the B/P bulge we computed the parameter $\rm B=(\sigma_z/\sigma_x)^2$
within the central $3$ kpc of the disk, where
$\sigma_z$ and $\sigma_x$ are the vertical and radial velocity dispersions
measured on a slit along the bar major axis. As discussed in \cite{Martinez06} (and references therein),
$\rm B\lsim 0.3$ corresponds to a buckling unstable galactic nucleus. As
expected, $\rm B$ decreases in time after the formation of the bar,
because of the rise of $\sigma_x$, down to the buckling unstable regime. As
soon as the B/P bulge forms $\rm B$ rises again, because of the increase
in  $\sigma_z$ associated with the buckling event. The buckling nature of the
B/P structure is still observable in the asymmetric (with respect to the
equatorial plane) mass distribution of the $z=0$ disk (see Figure~\ref{boxpean}). 
We note that $r_{\scriptsize{\mbox{A2}}}$ stops growing when the B/P structure
forms and  grows, consistently with the scenario of bars-weakening proposed
by, e.g., \cite{comb81, sell93, kor13}.



\section{Gas response to the bar growth and consequences on star formation}
\label{gas_response}

In this section we focus on the impact that the bar has on the evolution of the
gas and stellar component of the galaxy in the region dominated by the bar.

\subsection{Gravitational torque and gas evolution} \label{sec:torque}

\begin{figure*}
\centering
\includegraphics[width=0.45\textwidth]{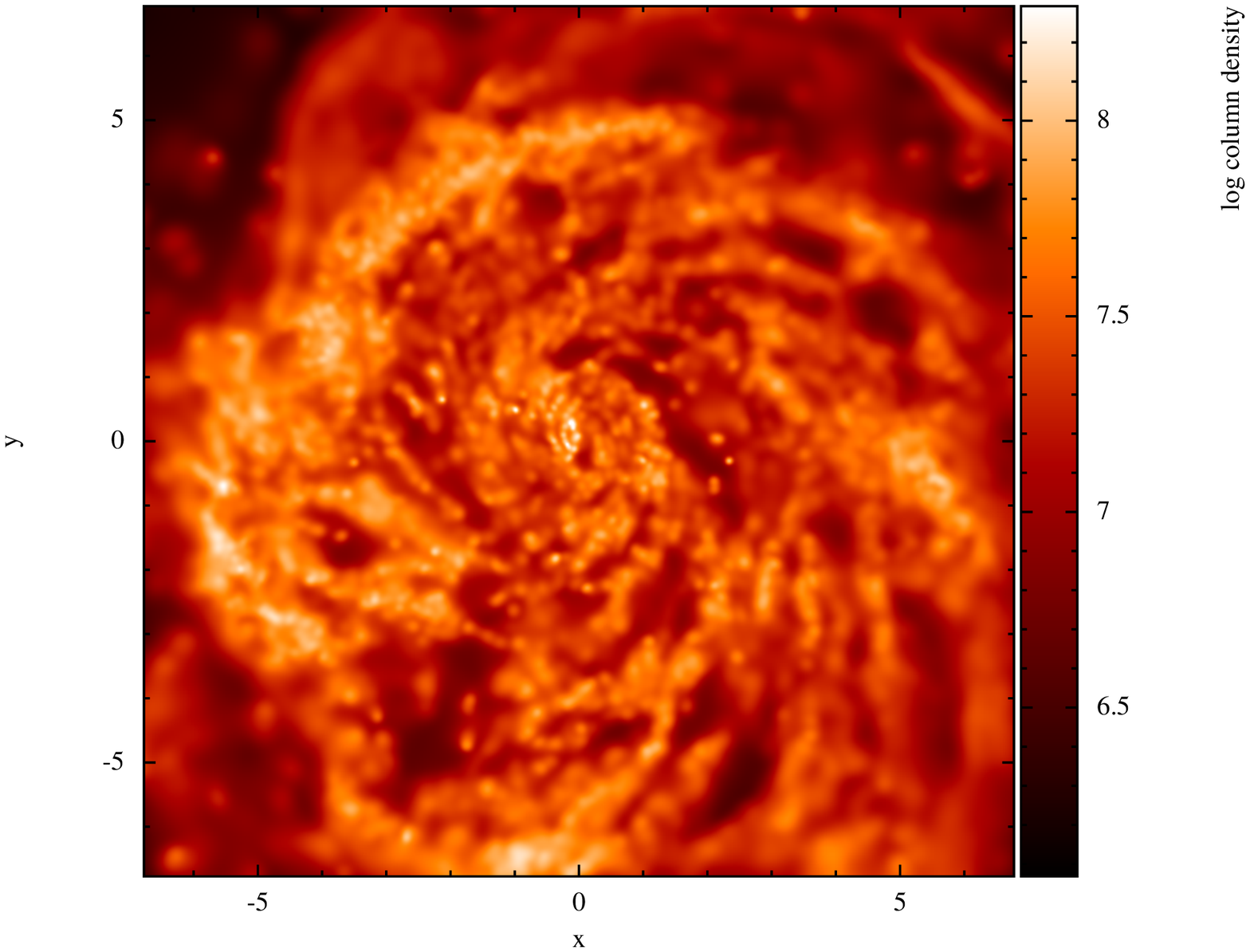}
\includegraphics[width=0.45\textwidth]{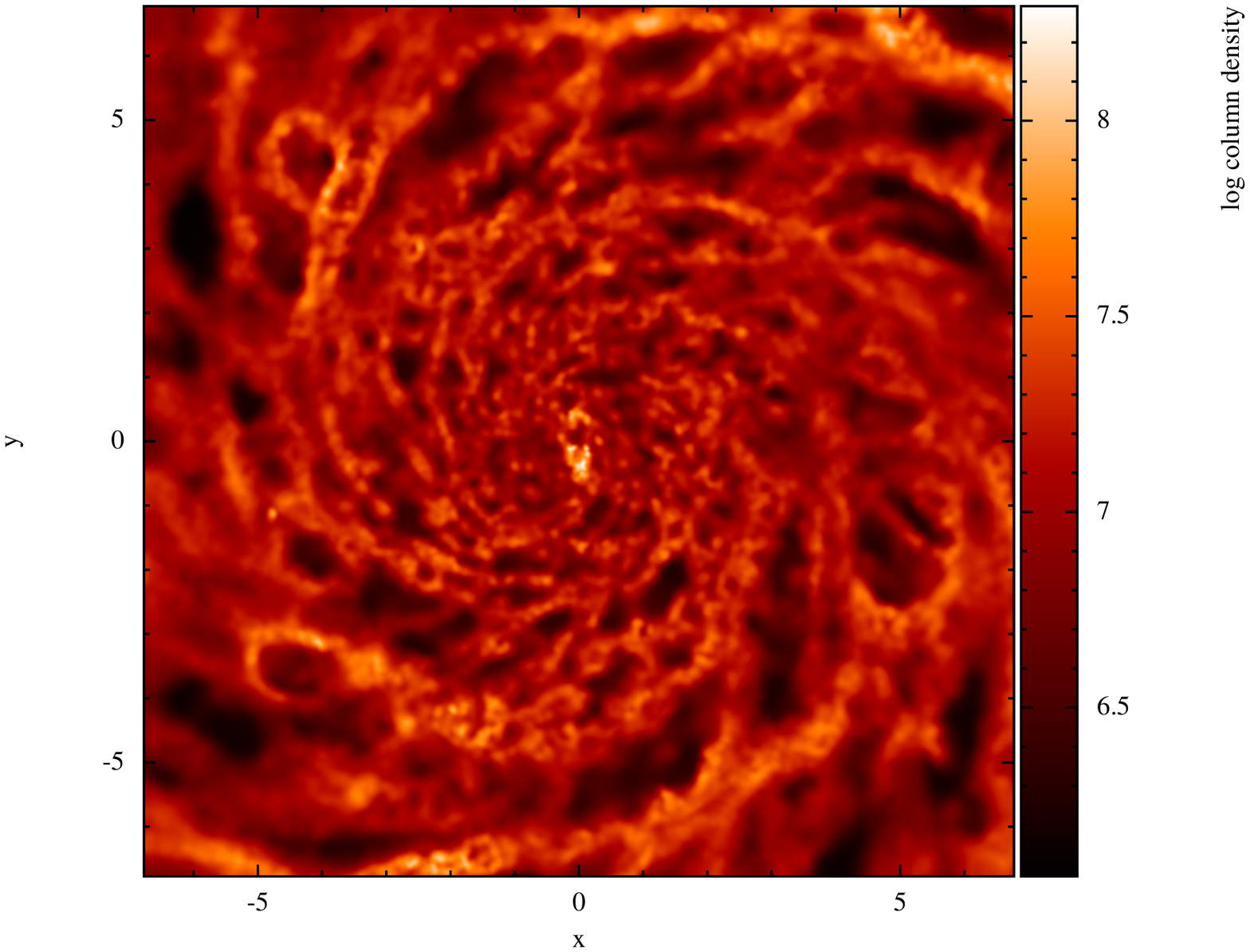}\\
\includegraphics[width=0.45\textwidth]{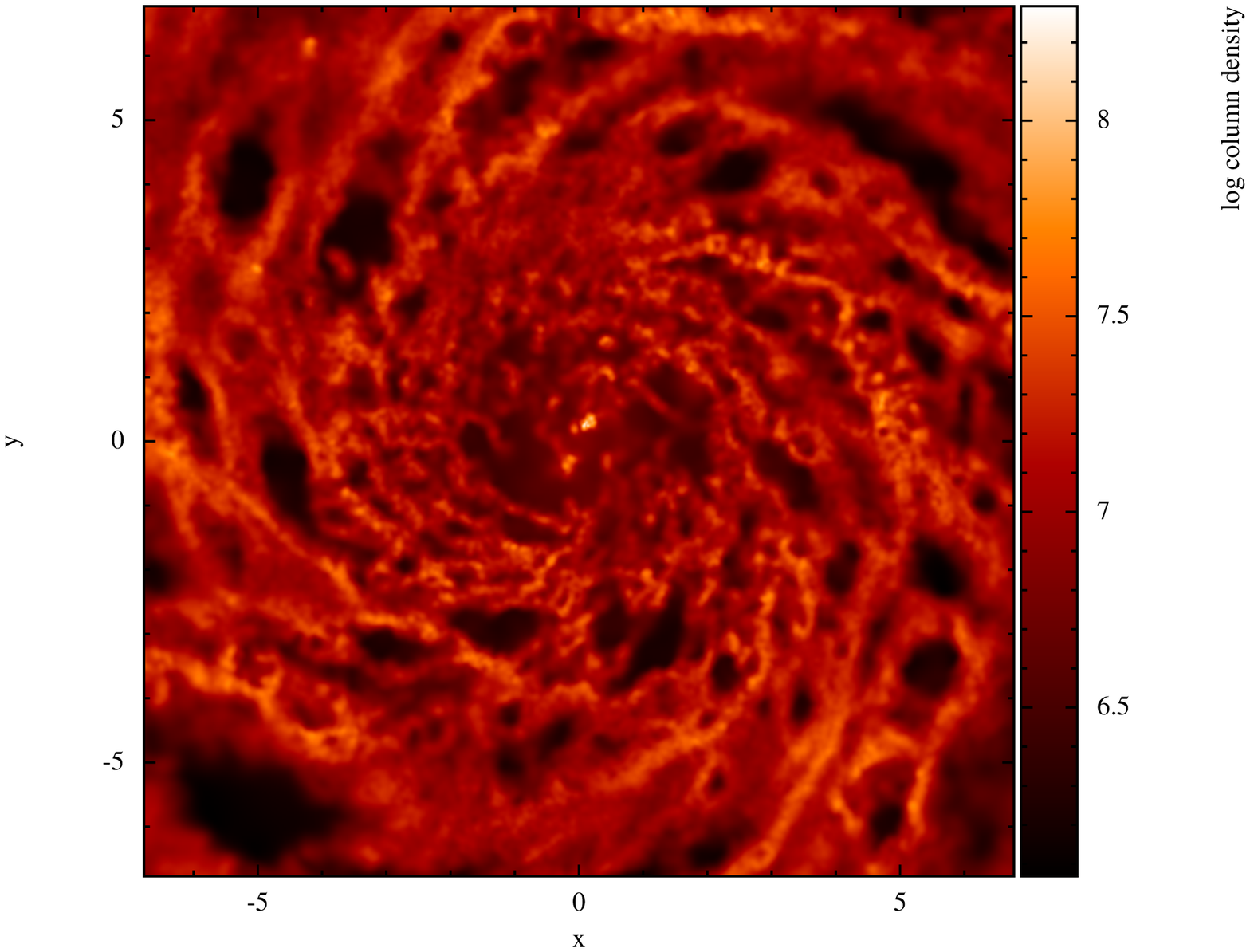}
\includegraphics[width=0.45\textwidth]{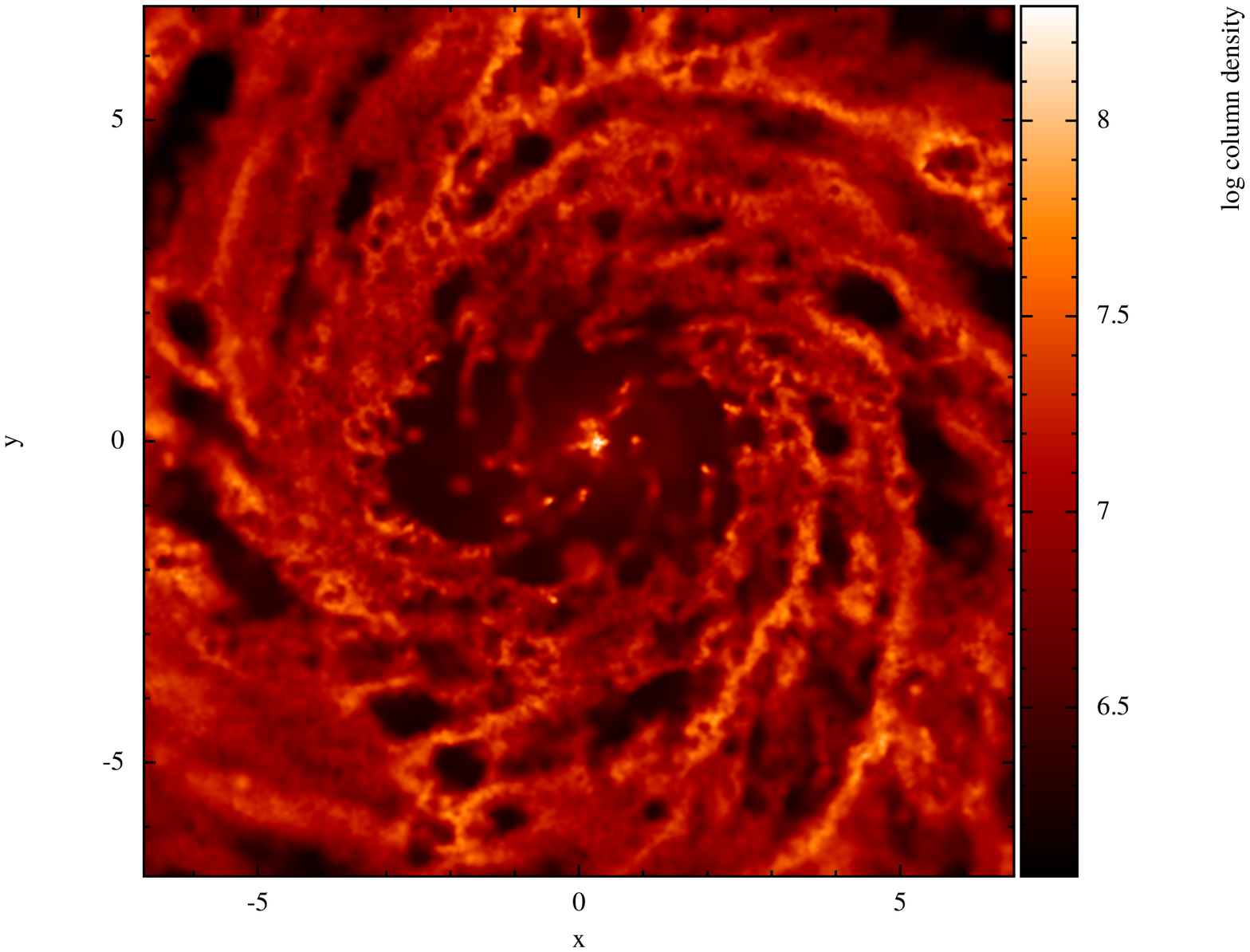}
\caption{Map of the gas surface density at four different epochs, $z=1.46$, $z=0.42$, $z=0.20$ and
  at the end of the simulation $z=0$ (upper left, upper right, lower left and
  lower right panel, respectively). Units are the same as in
  figures~\ref{render}, \ref{growth} and \ref{boxpean} but we plot the
   gas surface density in a red color-scale.}
\label{gasdensitymap}
\end{figure*}

\begin{figure}
\centering
\includegraphics[width=0.51\textwidth]{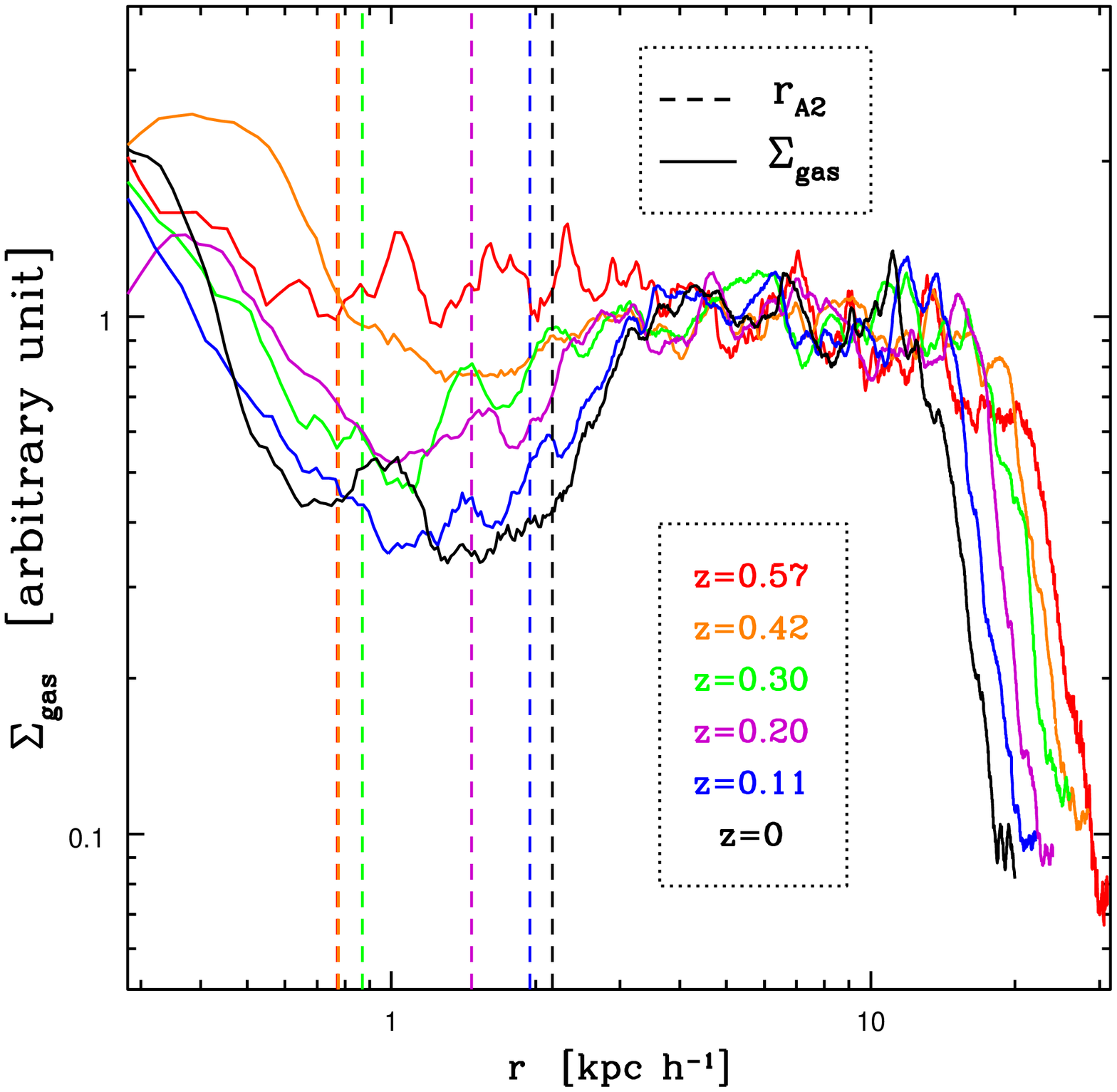}
\caption{Gas surface density profile at different times. The dashed vertical
  lines show the bar extent at the different times. Note that the surface
  densities has been renormalized to minimize the differences among the
  different profiles in the $4 \lsim R \lsim 10$ region.}
\label{gasdensityprofile}
\end{figure}

As the bar grows and gains strength, it starts exherting torques on the gas
component of the galaxy, modifying completely its distribution in the
central region. In 
Figure~\ref{gasdensitymap}  we show the surface density of the gas at four
different epochs.
The central region (approximately within $3$ kpc from the centre) of the galaxy
at $z=0$ appears almost empty of gas, except  for an unresolved density peak in
the galactic nucleus. The quantitative evolution of the gas content in the
galaxy center is shown in figure~\ref{gasdensityprofile}, where we show the
surface density profile of the gas at different times. In this case we renormalize the profiles in
the unperturbed region of the galaxy ($4\lsim R \lsim 10$ kpc). This allows to
emphasize the effect of the bar, averaging out the effects of cosmological gas
accretion and star formation-related gas consumption on large scales.

\begin{figure}
\centering
\includegraphics[width=0.44\textwidth]{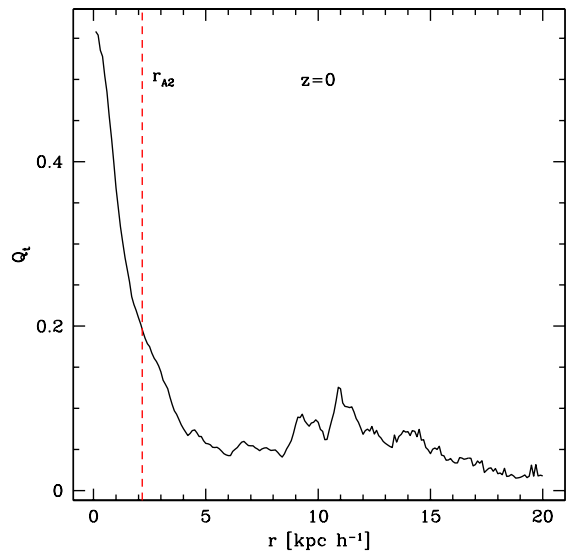}
\caption{Radial profile of $Q_{\mbox{t}}(r)$ at $z=0$ (black curve). The size of
the bar ($r_{\scriptsize{\mbox{A2}}}$) at the same redshift is indicated by the
red dashed line for reference. We see clearly a  high central value ($Q_{\scriptsize{\mbox{m}}}=0.56$ at $r\simeq0.15$ kpc) and no relative maximum at $r\simeq r_{\scriptsize{\mbox{A2}}}$. It is remarkable that the profile monotonically decreases up to $r\simeq4$ kpc from the centre. This shows that the bar non-axisymmetric structure is very coherent and the bar does not ``dissolve'' into a spherical bulge at small radii. This implies also the possibility that the bar efficiently drives the gas inflow up to the very central region of the disk.}
\label{torque}
\end{figure}

Figures~\ref{gasdensitymap}~and~\ref{gasdensityprofile} clearly show the
torquing effect that the growing bar has onto the gas. The gas within the bar
extent is driven towards the centre of the galaxy\footnote{We checked that the gas outflow from
the central region is negligible by calculating the total mass in star and gas
within the bar final extent ($r=2.17$ kpc) with respect to redshift during the bar growth phase.
We find that the
total baryonic mass is conserved
within $\sim3\%$ of its value at $z\sim0.45$ (before bar formation), thus excluding strong inflows/outflows of material.}, and the majority of it is
converted in stars (see below). 
Because of the absence of a clear ILR the gas
does not settle into a nuclear ring of star formation, but keeps on being
torqued by the bar down to the very central region of the galaxy, where it
forms an unresolved clump surrounded by a region completely depleated of gas
(see the bottom-right panel of Figure~\ref{gasdensitymap}). To confirm this
picture we estimate the relevance of the torque that the stellar distribution
exert onto the gas. Following \cite{mund99} and \cite{ems15} we calculate the
strength of the torque using: 
\begin{equation}
Q_{\mbox{t}}(r)=\frac{\small{\mbox{max}}\!\left[\frac{1}{r}\frac{\partial\phi(r;\theta)}{\partial\theta}\right]}{\left\langle\frac{\partial\phi(r;\theta)}{\partial r}\right\rangle_\theta}
\end{equation}
which, if, effectively, the ratio between the maximum tangential force and the mean axisymmetric
force at each radius $r$. The maximum value of $Q_{\mbox{t}}$ can also be used to
classify the bar strength, with $\mbox{max}(Q_{\mbox{t}})>0.4$ corresponding
to structures hosting strong bars \citep[e.g.][]{but04, but05}.  The torque
profile at $z=0$ is shown in Figure~\ref{torque}.
The maximum value of $Q_{\mbox{t}}$ is $Q_{\mbox{t}}\approx0.56$, confirming
the strong bar nature of the central non-axisymmetric structure. More
interestingly, the curve is peaked at very small radii close to our resolution
limit, which explains the formation of a compact central gas overdensity and
is consistent with the highly non-axisymmetric distribution of the stars at
the smallest radii (see Figure \ref{render}).
The bar does
  persists until $z=0$, thus most of the stellar mass in the inner 1-2 kpc
  remains associated with the bar rather than growing further the small
  pseudobulge. This is consistent with the notion that large central masses
  (of the order of a tenth of the total stellar disk) within a very compact
  size (well within one thenth of the disk scale length), are needed to
  destroy the bar \citep[see e.g.][]{shen04} while here the central
  overdensity is modest (about 3 \% of the total stellar mass within 300 pc),
  without any clear nuclear overdensity present.


We calculated the $Q_{\mbox{t}}(r)$ profiles at different times to sample the
bar strength evolution with respect to time. We find that the maximum torque
is always obtained near the galaxy centre (i.e. up to $r\simeq250$ pc),
confirming that the bar in \erisbh can be very effective in changing the gas
angular momentum up to the very smallest radii and make it fall towards the
centre.  This means that the bar can efficiently feed the central region,
providing the fuel necessary to ignite later evolutionary phenomena such as
nuclear star formation (e.g. \cite{kor13} and AGN activity
\citep[e.g.][]{comb00, quere15}.
In \erisbh, however, the accretion of matter onto the central MBH at
low z is very modest \citep{bonol15}. Most of the matter
  inflowing towards sub-kpc scales during the formation of the bar
  reaches densities large enough to be turned into stars (within a
  region of $\sim$ 600 pc), where nuclear star formation then becomes
the dominant process, as we discuss below.

\subsection{Star formation and black hole  accretion} \label{sec:sf}

\begin{figure}
\centering
\includegraphics[width=0.49\textwidth]{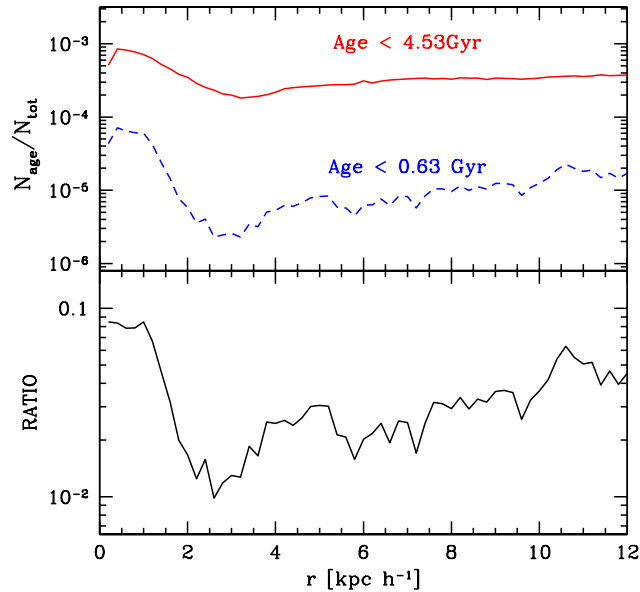}
\caption{Upper panel: radial distribution at $z=0$ of young stars formed after $z\approx0.05$ (i.e. with an age $<0.6$ Gyr, blue dashed line) and stars formed after the bar build-up at $z\approx0.4$ (with an age $<4.5$ Gyr, red solid line). The ratio of these two quantities (bottom panel) shows the signatures of a recent star formation episode (within $\sim1$ kpc) which transformed into stars the gas torqued down by the bar gravitational effect. This produced a gas-poor 
``dead zone'' between $2\lsim r\lsim3$ kpc (see also figure~\ref{gasdensitymap} bottom-right panel) where a low number of young stars is present.}
\label{SFprofile}
\end{figure} 

The strong central gas inflows caused by the torques exherted by the growing
bar, naturally lead to changes in the star formation and nuclear activity of
the galaxy. \cite{bonol15} already showed that the star formation rate and the
black hole accretion rate increase after $z \sim 0.2$, which is when the bar
 is reaching maximum strength (see Figure \ref{a2r2fig}). We further quantify
the effect of the gas inflow onto the central star formation and nuclear
activity here. In Figure~\ref{SFprofile} (upper panel) we show the radial
distributions of young stars (with an age $<0.6$ Gyr, i.e. formed after
$z\approx0.05$) and those formed after the build-up of the bar
structure at $z\approx0.4$ (i.e. those with an age $<4.5$ Gyr). The ratio
between these two quantities (bottom panel)
clearly points out the presence of a recent star formation episode in the very
central region of the galaxy (i.e. within $\approx1$ kpc). 

\begin{figure}
\centering
\includegraphics[width=0.49\textwidth]{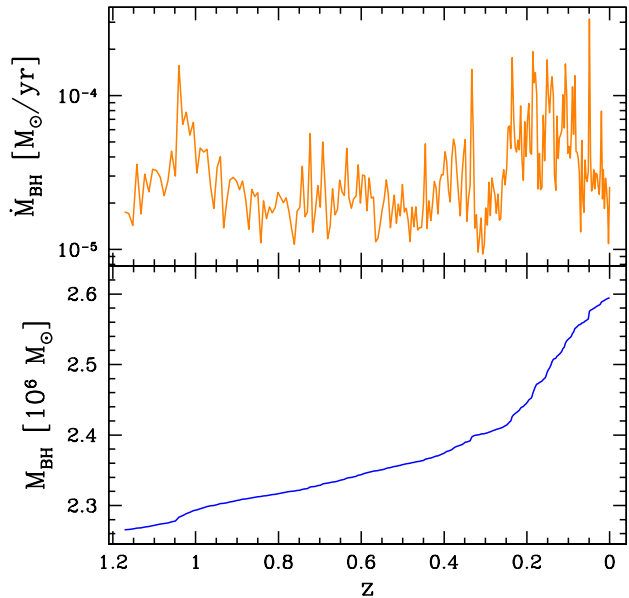}
\caption{Accretion rate (upper panel) and mass evolution (lower panel) of the
central black hole with respect to time. $\dot{M}_{\odot}$ is generally low
confirming that the BH mass growth by gas accretion is very small after the last
minor merger (about $\sim14\%$ of its final value). An increase in the MBH
accretion rate is observable during the development of the bar structure but
$\dot{M}_{\odot}$ remains modest even after $z\sim0.3$.}
\label{bhgrowth}
\end{figure}

A fraction of the inflowing gas gets accreted by the MBH. Figure
\ref{bhgrowth} (upper panel) shows the black hole accretion rate as a function
of redshift from the last minor merger to $z=0$.  $\dot{M}_{\odot}$ is
generally very low, fluctuating about $\rm 2\times10^{-5} M_\odot yr^{-1}$
with the exception of some isolated spikes \citep[see also][]{bonol15}. This
implies a modest growth of the black hole mass after $z\sim1.2$, which
undergoes a total increment of about $\sim14\%$ of its final value (lower
panel). A slight change in the accretion regime can be observed during the bar
growth phase for $z\lsim0.3$. The accretion rate results, however, in a
luminosity lower than $\sim 1\%$ of the MBH Eddington limit, assuming a
radiative efficiency of $\eta = 0.1$.  This further support the picture in
which the gas within the reach of the bar torques falls into the centre of the
galaxy and is promptly consumed by nuclear star formation bursts, while only a
very small fraction of it fuels the nuclear accretion process.
As the gas infall proceeds all the way to the center it leaves behind a low
  gas density region, a ``dead zone` visible in at 400 pc - 2 kpc in Figure 8,
  within which star formation can not be further sustained \citep{cheung13,
    gavazzi15, fanali15}. The bar in ErisBH does not extend out to its
  corotational radius (see Fig~\ref{growth}), i.e. its precession period is
  shorter than the orbital period of the outer gas. As a consequence the bar
  exerts a positive torque onto the outer gas, preventing any further gas
  infall that could potentially replenish the dead zone.

\begin{figure*}
\centering
\includegraphics[width=0.44\textwidth]{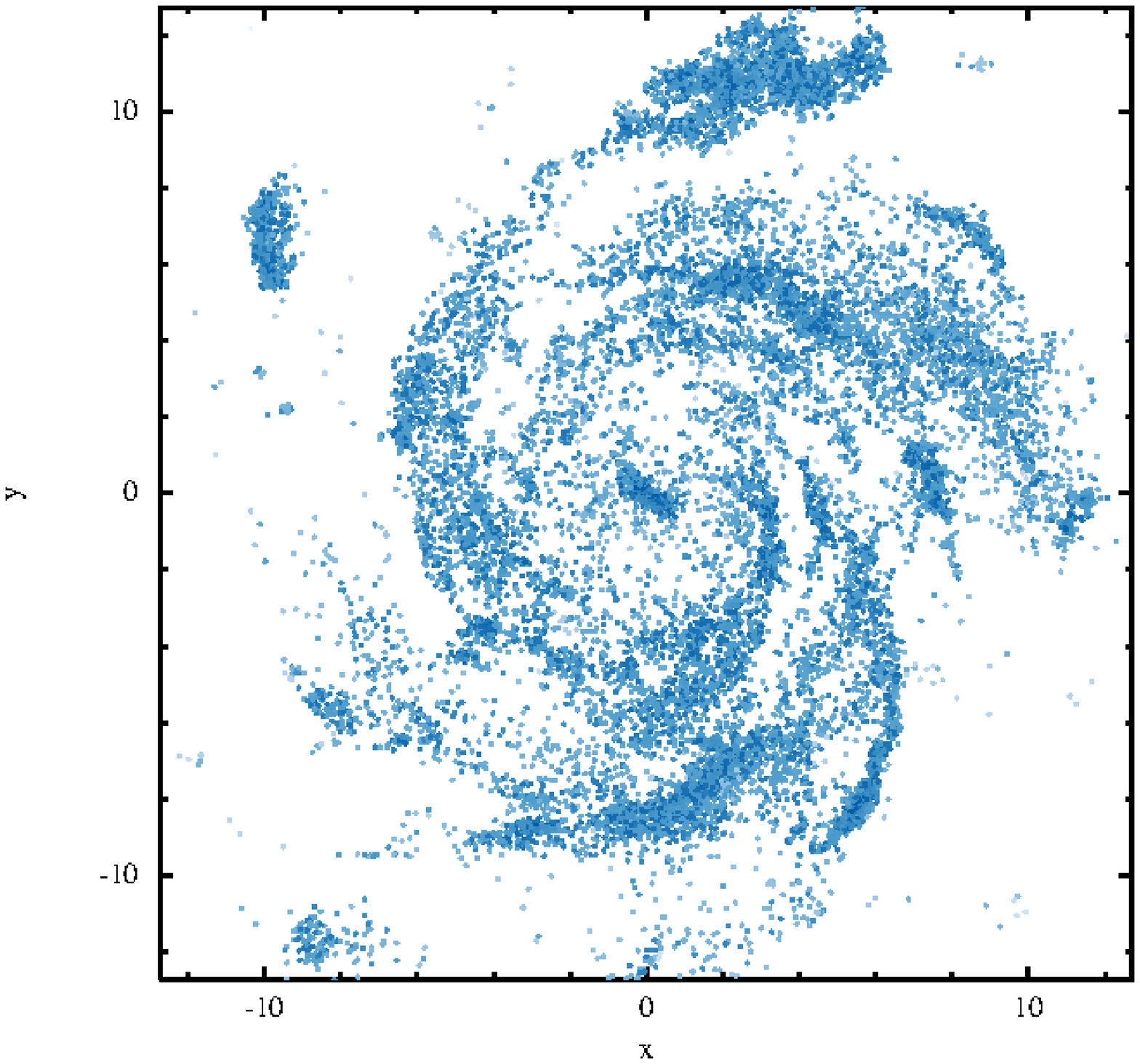}\qquad\quad
\includegraphics[width=0.41\textwidth]{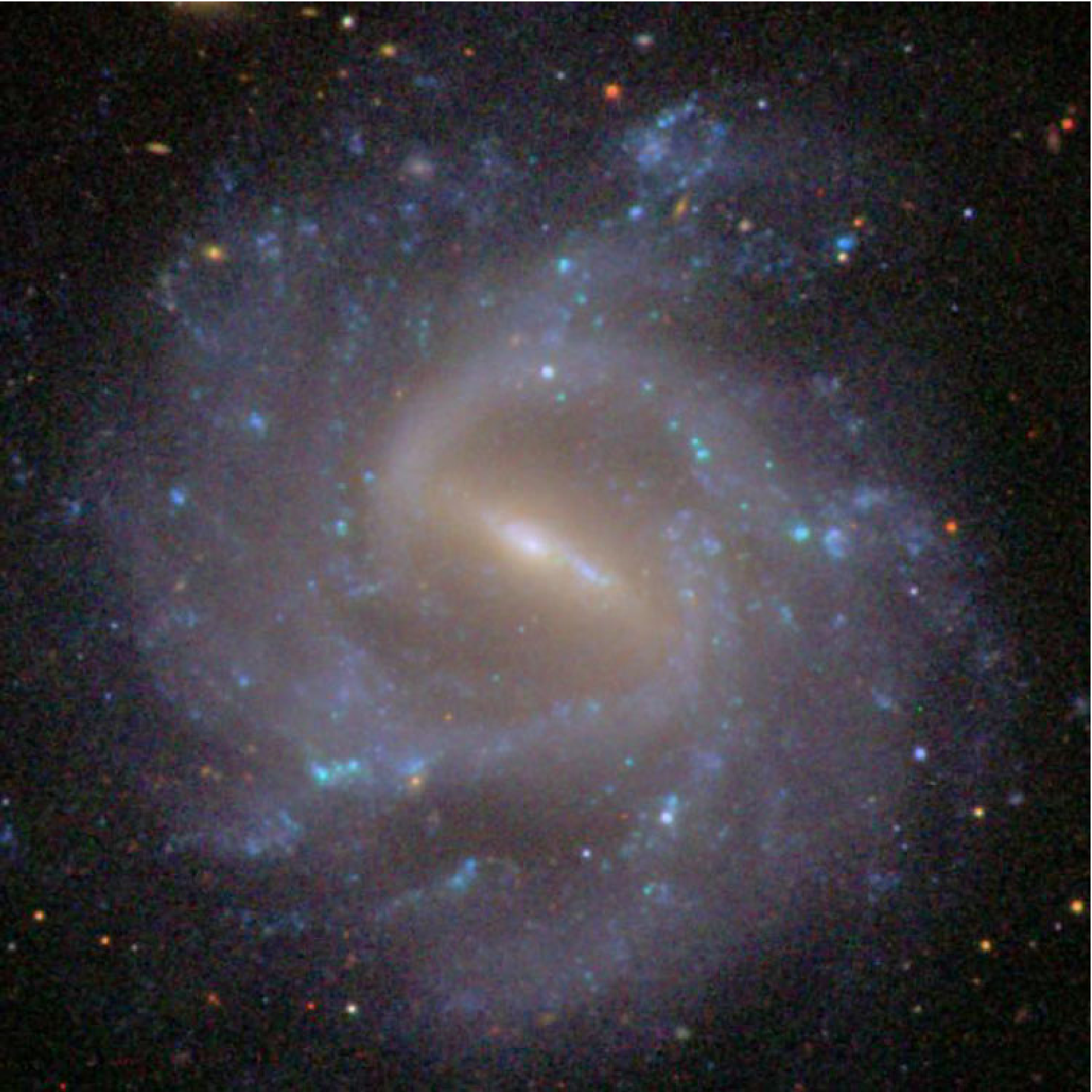}
\caption{Left panel: distribution at redshift $z=0$ of the youngest stars, i.e.
those with an age $<35$ Myr (formed at $z\approx 0.0026$, well after the bar
build-up). The outer regions of the disk are populated with young stars while a
``dead region'' in which few young stars are present is clearly visible in the
central region (i.e. up to $\sim2$ kpc from the centre). The dead region hosts a
central bar-like distribution of young stars. Right panel: the disk galaxy NGC
1073 is shown for a qualitative comparison. The outer regions of the NGC 1073
disk show star formation regions which host young stellar populations while the
inner regions exhibit a complex structure of older stars similar to that in
\erisbh.
}
\label{SFmap}
\end{figure*}

On the contrary, the formation of new stars proceeds unimpeded outside the
region affected by the bar. To further support this picture figure~\ref{SFmap}
shows the density weighted map of the stellar ages at the end of the
simulation. A population of young stars is clearly visible in the outer
regions of the galaxy, while only old stars are present in the dead region. A
nuclear ($\lsim 1$ kpc), elongated structure with intermediate age stars is
visible at the centre of the galaxy, in which the stars forming at the bar
onset contribute to the average age. A qualitative comparison with the similar
structure of NGC 1073 is shown in figure~\ref{SFmap}.  The outer disk in NGC
1073 is mostly composed of star formation regions which host young stellar
populations. On the contrary, a bar structure is evident in the galaxy centre
where the almost-exclusive presence of old and red stars is a prominent
feature.

\section{Conclusions}
\label{conclusions}

We analyzed the high-resolution cosmological \erisbh run \citep{bonol15},
which closely resembles an Sb/Sc galaxy with stellar mass and rotation velocity
comparable to our Milky Way.
At $z=0$ the galaxy forming at the centre of the refined region
features a strong nuclear ($R\approx 2$ kpc) bar which is able to strongly
influence:
$(i)$ the dynamics of the stellar disk, including the formation of a
B/P bulge in its centre; $(ii)$ the dynamics of the gas within the
central $3$ kpc, which falls towards the galactic centre triggering a
short burst of star formation in the galactic nucleus (within $\sim
600$ pc) as soon as the bar starts growing; $(iii)$ the late star
formation in the central $\sim 3$ kpc. This is the consequence of the
fast gas removal operated by the bar preventing any strong
star formation episode after its formation.

The analysis of the torques operated by the bar supports the notion that
the bar efficiently drives gas inflows down to the resolution limit ($\sim$ hundreds of
pc, due to the absence of any clear ILR) at any $z\lsim 0.4$. The absence of
an intense star formation activity in the central regions of the disk
as well as of strong AGN activity is purely due to the absence of dense gas
within the bar extent
due to rapid consumption by star formation at the onset of bar formation.
The lack of a clear
observational correlation between AGN activity and the occurrence of bars in
galaxies \citep[see e.g.][for the different point of views]{Ho97, Mulchaey97,
  Hunt99, Knapen00, Laine02, Lee12b, Alonso13, Cisternas13, Cheung15} could be
related to the prompt removal of gas.
If we assume the results of \erisbh appy to the whole class of
field disk galaxies in low density environments, we argue that the stronger 
gas inflow and enahnced star formation 
happen  at the onset of bar formation, when the detection
of a bar is more difficult as the bar is shorter and less regular in shape.
Instead, when the bar is stronger and well-developed, hence easily determined
from photometry or imaging, star formation has already ceased creating
a ``dead zone'' in the galactic centre and making the occurrence of
any nuclear activity less probable \citep[see e.g. the discussion
  in][]{fanali15}.

Strong bars may arise at earlier times in more massive galaxies or galaxies living
in dense environments, which evolve on shorter dynamical timescales.
Hence we argue that bar formation can contribute to quenching
and the formation of ``red nuggets'' at $z > 1$, as also suggested
by the results of the ARGO simulations which exhibit
several example of early bar formation leading to increased
central baryonic densities (Fiacconi, Feldmann \& Mayer 2015). Bar-driven
quenching should thus be seen as an alternative to mergers, disk fragmentation into massive clumps and
AGN feedback, the main mechanisms explored in the literature over the last few years.
Of course bar-driven quenching is related to feedback mechanisms operating in
the central region, as it seems to be the case in \erisbh where AGN feedback might
be instrumental in creating favourable conditions for bar formation at later stages.
Since bar-formation requires a kinematically cold, thin disk to occur, it remains
to be seen if this can be achieved by the latest generation of strong feedback
models adopted in galaxy formation simulations.

It is interesting to note that such a strong bar is absent in the Eris run,
which differs from \erisbh only because it does not feature any MBH accretion
and feedback prescription.
This would seem to be at odd with the limited gas accretion
occurring onto the central MBH \citep{bonol15}, that would imply a moderate
effect of AGN feedback onto the host galaxy. However, at $z > 1$ there are transient
near-Eddington accretion phases which ought to have an effect on the build-up
of the central baryonic distribution. Indeed at $z < 1$ \erisbh has a much flatter
rotation curve near the center as a result of the suppressed growth of the
central baryonic density.



The actual trigger of bar growth is still to be pinpointed. The main
galaxy in the \erisbh run becomes bar unstable at large redshift (see
Figure~\ref{growth}), but the bar structure forms only after the last minor merger episode. As
discussed in section~\ref{sfc_dnst}, the properties of the bar do resemble
those predicted for a tidally induced one.  Whether the merger itself does
provide the trigger for the instability to grow is unclear, as it is
impossible to definitively constrain the time in between the merger and the
actual onset of the bar growth. In order to test the possible tidal nature of
the bar we plan to run a set of simulations restarting the \erisbh run before
the merger, removing the particles forming the satellite, and checking whether
the bar grows regardless of the perturbation.

In conclusion, the present analysis of the \erisbh run has demonstrated that a
bar resulting from the fully cosmological evolution of a
disk galaxy with quiet merger history
strongly affects its host, in particular by quenching its star formation on
kpc scales. This result provides further theoretical support to the recent
claim by \cite{gavazzi15} that bars are one of the main contributor of the
flattening observed at high masses in the star formation rate-stellar mass
correlation \citep{whitaker12, magnelli14, whitaker14, gavazzi15b, ilbert15,
  lee15, schreiber16}.

\section*{Acknowledgments}
The Authors acknowledge Lia Athanassoula, Guido Consolandi, Victor de Battista, Luca Graziani \& Giuseppe Gavazzi for the insightful comments and
suggestions. Figures \ref{render}, \ref{growth} (right panels), \ref{boxpean}, \ref{gasdensitymap} and \ref{SFmap} (left panel)
were produced using the SPLASH visualization tool
	for SPH data \citep{price07}.

\bsp

\label{lastpage}

\end{document}